\documentclass[journal]{IEEEtran}
\IEEEoverridecommandlockouts
\usepackage{graphicx}
\usepackage{tabularx}
\usepackage{float} 
\usepackage[labelformat=simple]{subcaption}

\usepackage{algorithm}
\usepackage{algorithmic}
\usepackage{bm}

\usepackage{enumitem}
\usepackage{adjustbox}
\usepackage{multirow}
\usepackage{xcolor}
\usepackage{amsmath}
\usepackage{setspace}

\newenvironment{varalgorithm}[1]
{\algorithm[ht]\renewcommand{\thealgorithm}{#1}}
{\endalgorithm}

\usepackage{color,soul}
\setulcolor{red}
\setul{0.5ex}{0.3ex}
\usepackage{xcolor}

\newcounter{rtaskno}

\newcommand{\CASE}[1]{\STATE \textbf{case} #1\textbf{:} \begin{ALC@g}}
	\newcommand{\ENDCASE}{\end{ALC@g}}

\newcommand{\DEFAULT}{\STATE \textbf{default:} \begin{ALC@g}}
	\newcommand{\ENDDEFAULT}{\end{ALC@g}}
\newcommand{\DEFAULTLINE}[1]{\STATE \textbf{default:} }

\usepackage{newfloat}
\DeclareFloatingEnvironment[placement={!ht},name=List]{mylist}

\def\BibTeX{{\rm B\kern-.05em{\sc i\kern-.025em b}\kern-.08em
		T\kern-.1667em\lower.7ex\hbox{E}\kern-.125emX}}

\usepackage{listings}
\definecolor{mygray}{rgb}{0.5,0.5,0.5}

\begin{document}
	
	\thispagestyle{plain}
	\pagestyle{plain}
	
	\title{A Novel ASIC Design Flow using Weight-Tunable Binary Neurons as Standard Cells}
	\author{Ankit Wagle, Gian Singh,~\IEEEmembership{Member,~IEEE}, Sunil Khatri,~\IEEEmembership{Senior Member,~IEEE}, Sarma Vrudhula,~\IEEEmembership{Fellow,~IEEE}
		\thanks{Authors A. Wagle, G. Singh, and S. Vrudhula are with the School of Computing and Augmented Intelligence, Arizona State University, Tempe, AZ, USA (e-mail: (awagle1,gsingh58,vrudhula)@asu.edu)}
		\thanks{Author S. Khatri is with the Dept. of Electrical and Computer Engineering, Texas A\&M University, College Station TX, USA (e-mail: sunilkhatri@tamu.edu)}
		\thanks{The research was supported in part by NSF PFI award \#1701241, \#1361926.}
		\thanks{Copyright~\copyright~2022 IEEE. Personal use of this material is permitted. However, permission to use this material for any other purposes must be obtained from the IEEE by sending a request to pubs-permissions@ieee.org.}
		\vspace{-20pt}
	}
	
	\maketitle

	\begin{abstract}
		In this paper, we describe a design of a mixed-signal circuit for a binary neuron (a.k.a perceptron, threshold logic gate) and a methodology for automatically embedding such cells in ASICs.  The binary neuron, referred to as an FTL (flash threshold logic) uses floating gate or flash transistors whose threshold voltages serve as a proxy for the weights of the neuron.  Algorithms for mapping the weights to the flash  transistor threshold voltages are presented. The threshold voltages are determined to maximize both the robustness of the cell and its speed. The performance, power, and area of a single FTL cell are shown to be significantly smaller (79.4\%), consume less power (61.6\%), and operate faster  (40.3\%) compared to conventional CMOS logic equivalents.   Also included are the architecture and the algorithms to program the flash devices of an FTL.  The FTL cells are implemented as standard cells, and are designed to allow commercial synthesis and P\&R tools to automatically use them in synthesis of ASICs.  Substantial reductions in area and power without sacrificing performance are demonstrated on several ASIC benchmarks by the automatic embedding of FTL cells.    The paper also demonstrates how FTL cells can be used for fixing timing errors after fabrication.
	\end{abstract}
	
	
	\section{Introduction and Motivation}
	\label{sec:intro}
	
	In this paper, we introduce a new \textit{programmable ASIC primitive}, referred to as a \textit{flash threshold logic} (FTL) cell, and show how it can be used to substantially improve the area and power consumption of an ASIC, without increasing its delay.  An FTL cell and its use in an ASIC is different from any other type of ASIC component previously reported.  It is a mixed-signal circuit that is designed as a \textit{standard cell}, so that it is fully compatible with conventional CMOS  logic synthesis, technology mapping, and place-and-route tools.  
	
	An FTL cell of $n$ inputs realizes any \textit{threshold} function of $n$ or fewer variables. A threshold function $f(x_1, \cdots, x_n)$~\cite{MurogaBook71} is a unate Boolean function whose on-set and off-set are \textit{linearly separable}, i.e. there exists a vector of weights $\bm{W} = (w_1, w_2, \cdots, w_n)$\footnote{W.L.O.G, weights can be assumed to be positive integers~\cite{Siu_1995}, and for a given truth table of a threshold function, there is a weight vector whose sum is minimum~\cite{Siu_1995}.} and a threshold $T$ such that
	
	\begin{equation}
		\label{eq:ThresholdDefinition}
		f(x_1, x_2, \cdots, x_n) = 1 \Leftrightarrow \sum_{i=1}^{n} w_i x_i \geq T,
	\end{equation}
	where $\sum$ here denotes the arithmetic sum.  A threshold function can be equivalently represented by $(\bm{W}, T) = (w_1, w_2, \cdots, w_n; T)$.
	
	\begin{figure}[h]
		\centering
		\includegraphics[width=\columnwidth]{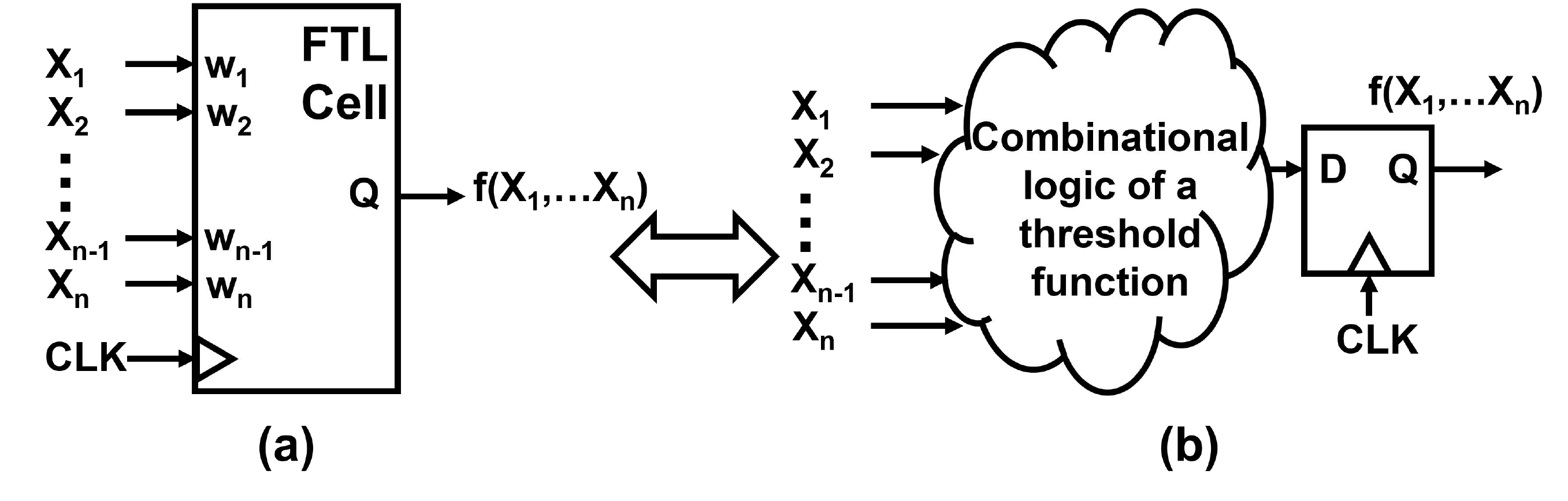}
		\caption{(a) FTL Schematic, (b) Functional equivalent}
		\label{fig:FTLSchematic}
		\vspace{-10pt}
	\end{figure}
	
	Figure~\ref{fig:FTLSchematic}(a) shows a block diagram of an FTL cell, in which the weights $\bm{W}$ are shown as internal parameters of the cell and Figure~\ref{fig:FTLSchematic}(b) shows its functional equivalent.  The schematic is meant to convey that the input-output behavior of an FTL cell may be viewed as an  \textit{edge-triggered}, \textit{multi-input} flip-flop, whose output is the value of a threshold function, that is internally latched at the rising edge of the clock signal CLK.  
	
	A distinctive characteristic of the FTL cell design is that the actual threshold function realized by an FTL instance within an ASIC is \textit{programmed after the circuit is manufactured}.  An FTL-based ASIC integrates \textit{flash} or \textit{floating gate}~\cite{Cai_DATE_2013} transistors along with conventional MOSFETs within the FTL cell.  Thus, unlike many of the \textit{emerging technologies}~\cite{Perricone_DATE_2017,Yang_NANOARCH_2014,jha_nanopipelining,berezowski:dsd05}, an FTL cell employs mature IC technologies (CMOS and Flash) that are routinely integrated today and commercially manufactured with high yields.
	
	\subsection{FTL in ASIC Design~--~Overview}
	\label{subsec:ASICDesignUseCase}
	
	\begin{figure*}%
		\centering
		\begin{subfigure}{.3\textwidth}
			\centering
			\includegraphics[scale=0.35]{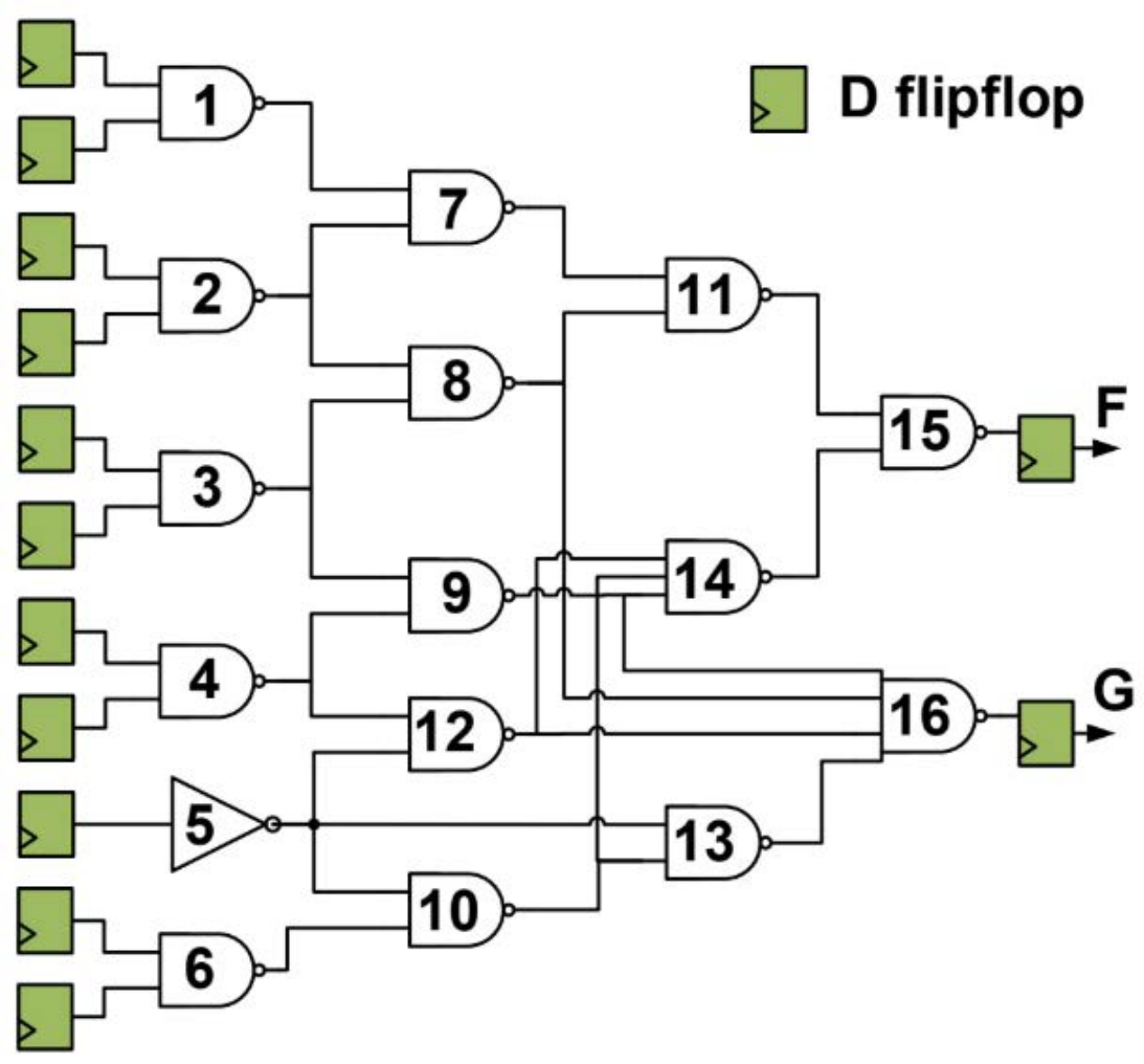}%
			\caption{A logic netlist.}%
			\label{subfiga:FTL_ASIC}%
		\end{subfigure}\hfill%
		\hspace{-30pt}
		\begin{subfigure}{.30\textwidth}
			\centering
			\includegraphics[scale=0.35]{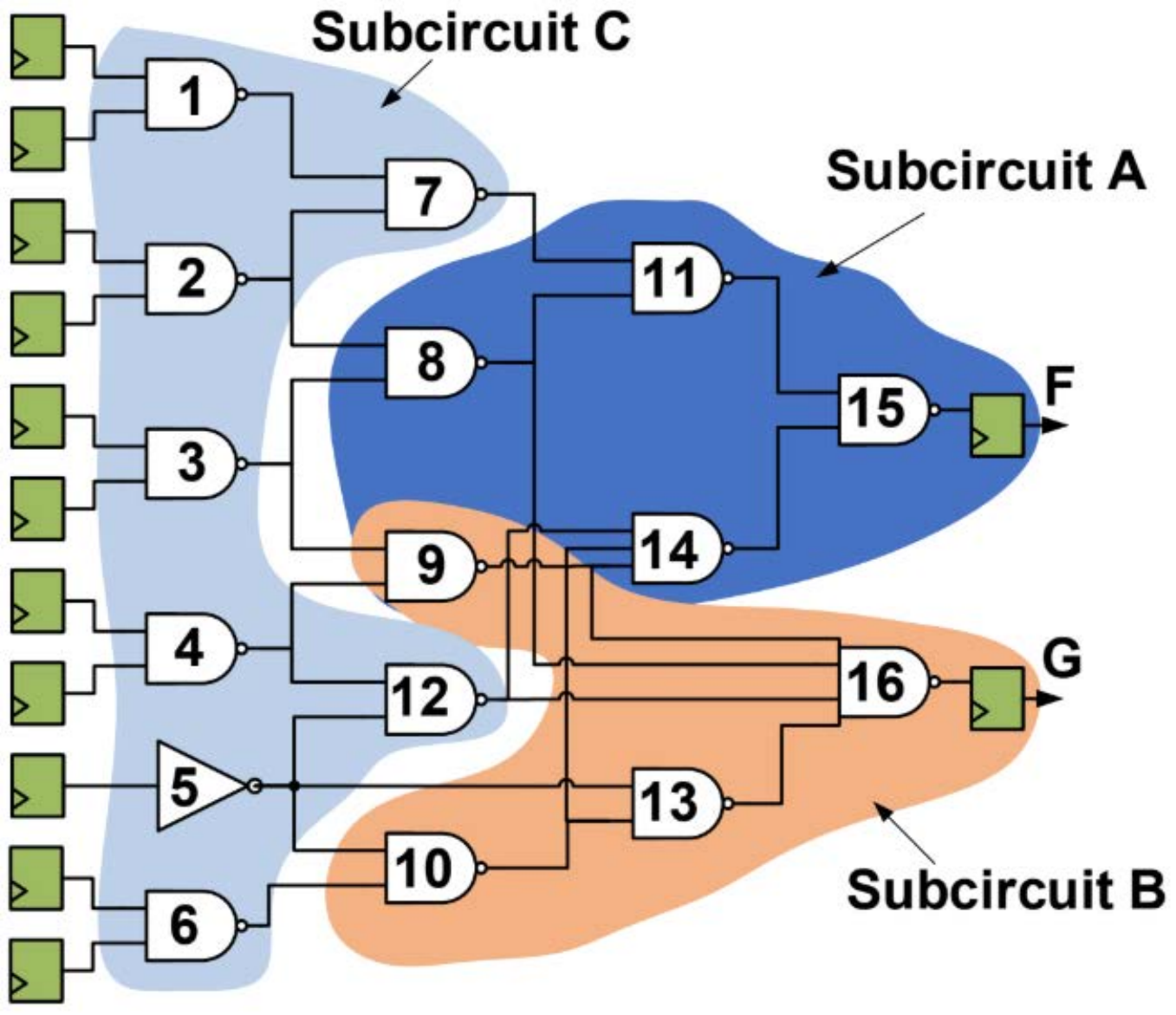}%
			\vspace{-5pt}
			\caption{Identifying threshold functions in TFI cones of flip-flops}%
			\label{subfigb:FTL_ASIC}%
		\end{subfigure}\hfill%
		\begin{subfigure}{.3\textwidth}
			\centering
			\includegraphics[scale=0.37]{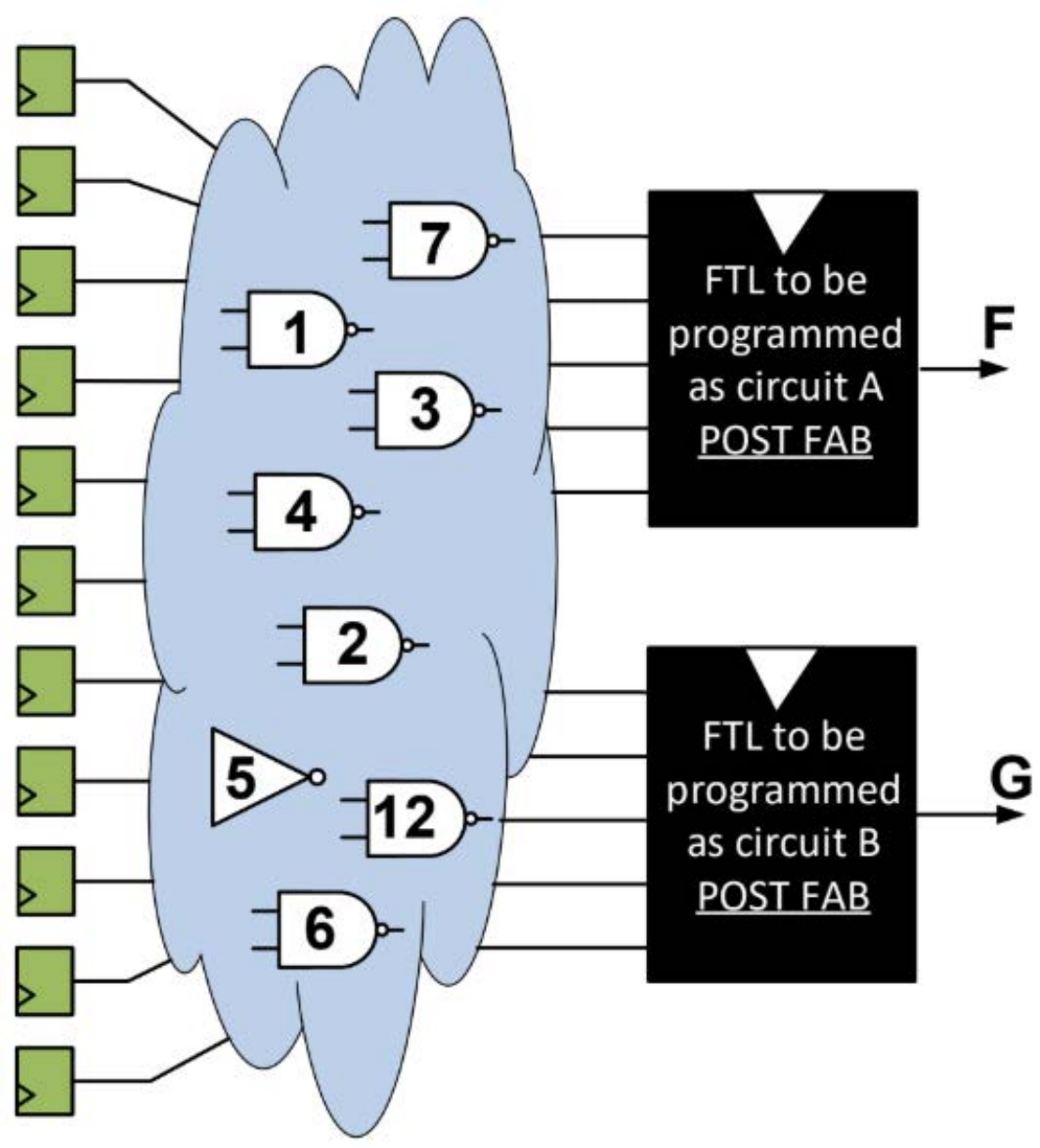}%
			\caption{An FTL-CMOS logic hybrid}%
			\label{subfigc:FTL_ASIC}%
		\end{subfigure}%
		\caption{Use of FTL in ASIC design.}
		\label{fig:FTL_ASIC}
		\vspace*{-10pt}
	\end{figure*}
	
	Before proceeding to the details of FTL design, it will be instructive to understand how it can be used in an ASIC~\cite{Kulkarni_TVLSI_2016}, and how it can improve performance, power, and area. 
	
	Consider the logic netlist shown in Figure~\ref{subfiga:FTL_ASIC} which has two registered outputs $F$ and $G$. Suppose that the transitive fan-in (TFI) cones of $F$ and $G$ are traversed and two subcircuits labeled $A$ and $B$ (see Figure~\ref{subfigb:FTL_ASIC}) are found, such that $A$ and $B$ are threshold functions of their inputs. The remaining subcircuit is labeled as $C$.  Suppose that subcircuits $A$ and $B$ are replaced by FTL cells, which are to be later programmed to realize $A$ and $B$.  This replacement is shown in Figure~\ref{subfigc:FTL_ASIC}, where the FTL cells are shown as black boxes.  Now, subcircuit $C$ would be re-synthesized to account for the changes in the delay of FTL cells and the new loads that the inputs of the FTL cells present to the outputs of $C$.  There are two reasons why the circuit in Figure~\ref{subfigc:FTL_ASIC} would have improved power, performance and area. 
	
	\begin{enumerate}
		\item Subcircuits $A$ and $B$ and the two flip-flops are each replaced by an FTL cell which has much fewer transistors, resulting in a significant reduction in area and power.
		
		\item  The clock-to-Q delay of FTL cells turns out to be much less than the total delay (combinational logic delay plus clock-to-Q delay of DFF) of subcircuits $A$ and $B$, which results in creating a substantial amount of slack (required time minus arrival time) on the outputs of subcircuit $C$. This in turn will allow synthesis and technology mapping tools to reduce the logic area of subcircuit $C$.  The extent of the improvement depends on how the logic is \textit{absorbed} into the FTL cell. 
	\end{enumerate}
	
	Note that the reason why the FTL cells are shown as black boxes in Figure~\ref{subfigc:FTL_ASIC} is to convey the fact that their functions are not known at the time of fabrication because the flash transistors are \textit{programmed} after the chip is manufactured.  
	
	\subsection{Main Contributions}
	\begin{enumerate}
		\item An FTL cell is a mixed-signal circuit that is implemented as a standard cell. The new design incorporates flash transistors, which allow its function to be \textit{programmed} after fabrication.
		
		\item The set of threshold voltages of the flash transistors in an FTL cell serve as a proxy for the weights $[\bm{W}, T]$.  The weights can be realized with great fidelity because the flash transistors can be programmed with high precision~\cite{Cai_DATE_2013}.  However, the relationship between the weights and threshold voltages is a non-linear and multi-valued mapping that depends on the complex electrical and layout characteristics of the MOSFETs and flash transistors.  We present a new algorithm called the \textit{modified perceptron learning algorithm} (mPLA)~\cite{rosenblatt_PR_1958} that works in concert with HSPICE and \textit{learns} a mapping between the weights and threshold voltages.  The mPLA algorithm also maximizes the noise tolerance and robustness of the FTL cell in the presence of process and environmental variations.
		
		\item We present an efficient architecture and methodology for programming the threshold voltages of each flash transistor within each FTL cell that is embedded in an ASIC.  We also demonstrate how the post-fabrication threshold voltage assignment capability can be used to improve functional yield and correct timing errors.
		
		\item FTL cells are designed as standard cells to be compatible with existing CMOS design methodologies and tools.  We demonstrate this compatibility by using commercial CAD tools to perform synthesis,  technology mapping, and place-and-route of several complex function blocks with FTL cells included in the cell library. The results show that automatic embedding of FTL cells results in substantial improvements in the area, power, and wirelength, without sacrificing performance.
	\end{enumerate}
	
	The remainder of the paper is organized as follows. Section~\ref{sec:background} gives a very brief overview of threshold logic and flash transistor technology. Sections \ref{sec:FTLArchitecture}, \ref{sec:ftl_asic_integration}, \ref{sec:WeightsToVTs} and \ref{sec:ExpResults} contain the main body of this work.  The architecture and operation of the FTL cell are described in Section~\ref{sec:FTLArchitecture}.  Section~\ref{sec:ftl_asic_integration} explains the mechanism for programming FTL cells once they are embedded in an ASIC, using a separate scan chain reserved for that purpose.  Section~~\ref{sec:WeightsToVTs} describes the mapping between the weights of a threshold function and the threshold voltages of the flash transistors in the FTL, considering factors such as robustness to noise, process variations and circuit delay. Section~\ref{sec:ExpResults} contains an extensive set of experimental results, demonstrating the significant improvements in PPA of FTL cells over their CMOS equivalents both at cell-level and when they are integrated into ASICs.  It also includes results validating several uses of post-fabrication programming/tuning of the flash devices.  Conclusions appear in Section~\ref{sec:Conclusions}.  
	
	\section{Background}
	\label{sec:background}
	
	\subsection{Threshold Logic}
	
	Threshold functions are an interesting and valuable subset of Boolean functions. They were first proposed in 1943 as simple models of neurons~\cite{McCulloch:1988}, which generated a vast number of papers on neural networks~--~a subject that has been revived recently with the emergence of machine learning.  The use of threshold logic in digital design and synthesis was extensively investigated in the 1960s and 1970s, culminating in two authoritative works~\cite{DertouzousBook:1965,MurogaBook71}.  Since then there has been a substantial body of work on new circuit architectures and implementations of threshold logic.  Surveys of designs prior to 2003 appear in~\cite{Beiu_2003_TNN,Beiu_IJCNN_2003,Celinski_2003}, detailing over fifty different implementations.
	
	The earlier works and even later ones such as ~\cite{Lageweg:ICECS2002, Mozaffari_TCSI_2018,zhang:tcad2005,ANNAMPEDU201384,Mozaffari_TNANO_2018, Savas_SSE_2007, Yang_NANOARCH_2014},  have not been integrated into mainstream VLSI design. It is, however, still is very valuable to develop efficient implementations of threshold functions. This is due to the fact that many Boolean functions that require large AND/OR networks can be realized by smaller, fixed depth threshold networks~\cite{Siu_1995} and nearly 70\% of the functions in two standard cell libraries (observed in a 65nm and a 40nm library from different vendors) are threshold functions. 
	
	Recently, \cite{Kulkarni_TVLSI_2016} reported an architecture of a threshold gate and showed how it can be integrated with the standard-cell ASIC design methodology using commercial tools. Unlike our approach, \cite{Kulkarni_TVLSI_2016} uses only CMOS devices. In addition, they reported significant improvements in PPA of an actual silicon implementation of ASIC with threshold gates~\cite{Jinghua:CICC2015}.  Their architecture, however, severely limits the number of threshold functions that can be implemented because the width of the transistors are made proportional to the weights.  This limits the fan-in and consequently, only 12 of the 117 threshold functions of 5 or fewer inputs were implemented in~\cite{Kulkarni_TVLSI_2016}. In contrast, our work implements all 117 threshold functions of 5 or fewer inputs because of the use of flash transistors.
	
	\subsection{Flash Transistors}
	\label{subsec:background:flash}
	
	A flash transistor is functionally similar to a field effect (FET) transistor, except that it is made to have an additional layer of charge between the control gate and the channel as a means to adjust the threshold voltage ($VT$) of the transistor. Programming and erasing these devices correspond to increasing and decreasing $VT$, and this is achieved by electrons tunneling into or out of the charge layer via Fowler-Nordheim (FN) tunneling~\cite{Fowler_RSL_1928}.  In general, programming is performed by applying a sequence of high voltage pulses (the duration and magnitude of which determines $VT$)~\cite{Richter2014} to the control gate and holding the bulk, source, and drain terminals at $0V$. Erasure is achieved by holding the control gate at $0V$ and allowing the source and drain to float~\cite{Richter2014}. In a flash memory, the value of $VT$, which is determined by sensing the current, represents the state or stored value, and this can be retained for more than ten years~\cite{Richter2014}, while the bulk is driven to a high voltage. Several variants of flash transistors with different structures and materials have been investigated over the past two decades to reduce the programming voltage, improve reliability, encode multiple bits, reduce the number of fabrication steps, and improve the yield.
	
	\begin{figure}[ht!]
		\vspace{-10pt}
		\centering
		\begin{subfigure}{.20\textwidth}
			\includegraphics[width=\columnwidth]{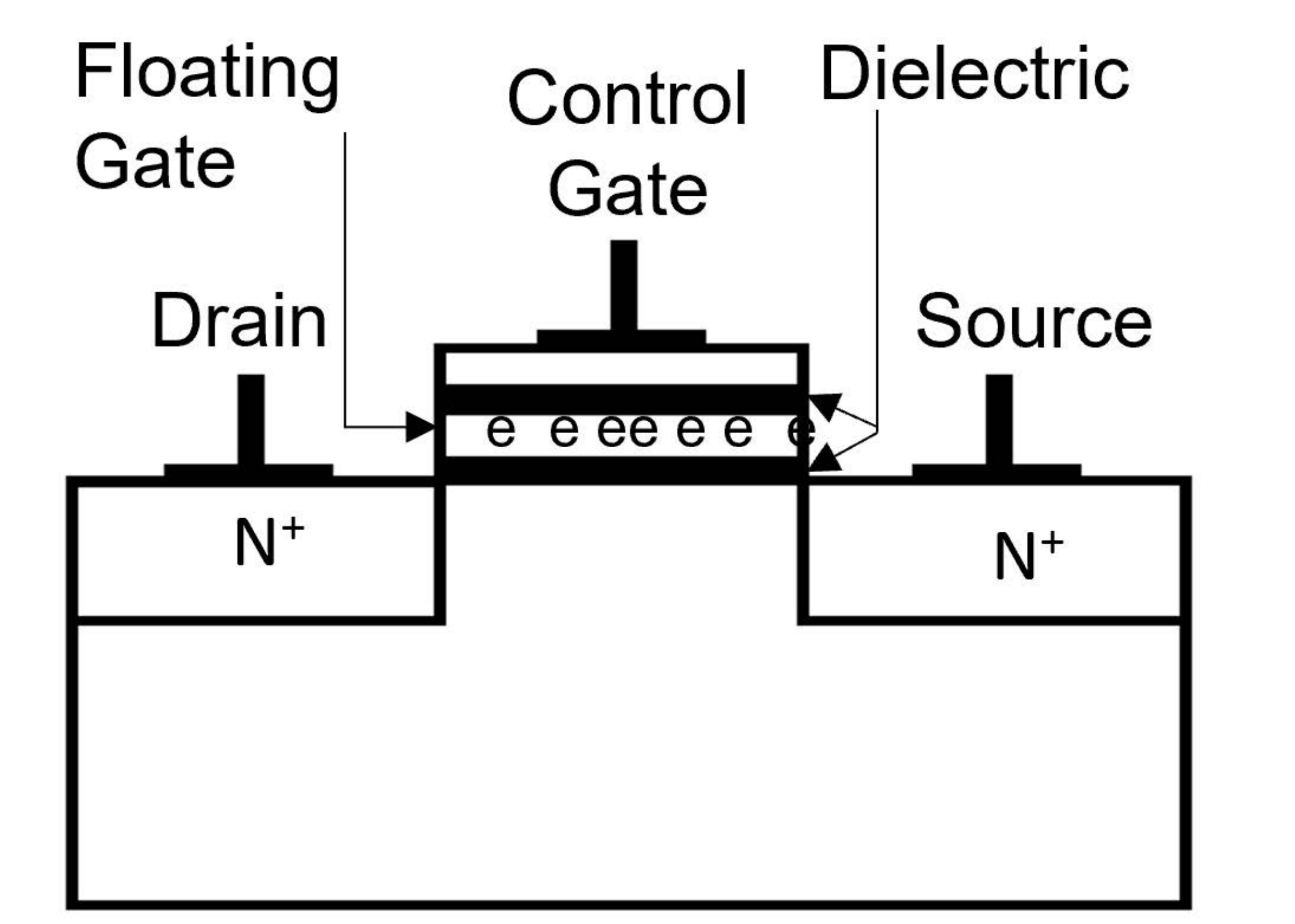}
			\caption{Floating gate Transistor}%
			\label{fig:FGT}
		\end{subfigure}
		\begin{subfigure}{.21\textwidth}
			\includegraphics[width=\columnwidth]{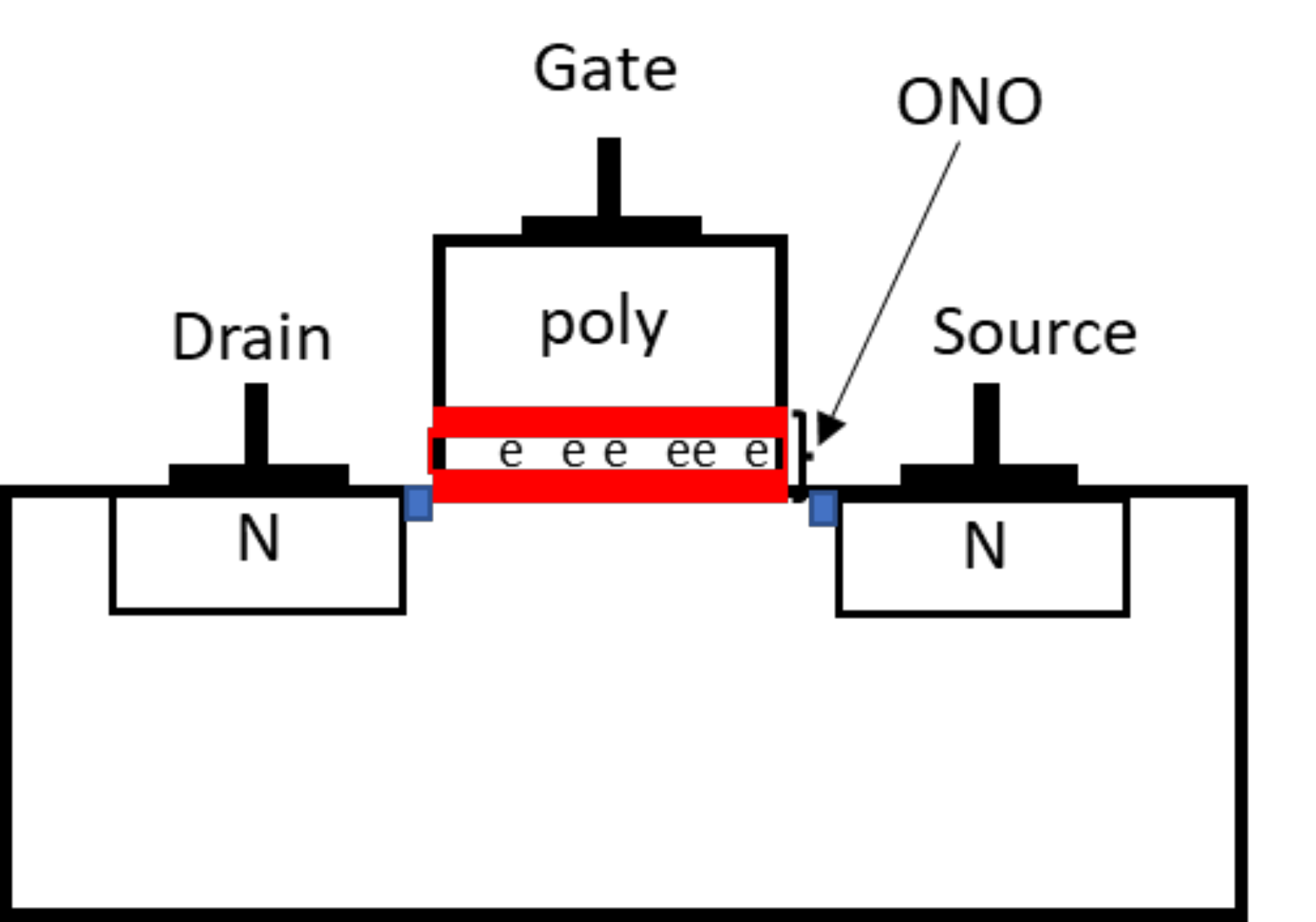}
			\caption{Charge Trap Transistor}
			\label{fig:CTT}
		\end{subfigure}
		\vspace{-5pt}
		\caption{\label{fig:flash_cross_section} Cross section of flash transistors.}
		\vspace{-5pt}
	\end{figure}
	
	Figure~\ref{fig:flash_cross_section} shows a cross-section of the two common types of flash transistors: (a) the floating gate transistor (FGT), which has an additional buried and un-contacted floating gate, and (b) the charge trapping transistor (CTT) which has an oxide-nitride-oxide (ONO) layer between the gate and the substrate.  Floating gate technology (Figure~\ref{fig:FGT}) is fully compatible with and used along with CMOS, and because of its dominant role as NVM in flash drives and solid-state drives (SSD), its design and fabrication has been continuously improved over two decades.  However, it still has several drawbacks, including the need for additional masks, the requirement of higher voltages for programming and erasure, and most importantly, the difficulty in scaling its dimensions below 40nm due to poor scaling of the thin oxide.
	
	In CTTs (Figure~\ref{fig:CTT}), the electrons are trapped in the insulating nitride layer whereas in FGTs they are in the conducting floating gate.  There are several variants of CTT device such as SONOS~\cite{nii_cost-effective_2020}, MONOS\cite{tsuda_first_2016} and High-K Metal Gate (HKMG) \cite{khan_turning_2019}.  All have been successfully scaled to 14nm/16nm CMOS FinFET technology. Additionally, the HKMG device can be programmed to multiple $VT$ levels \cite{khan_charge_2017} and the SONOS device can be programmed up to 128 $VT$ levels \cite{agrawal_memory_2020}.  One important advantage of the HKMG CTTs is that they do not require any additional processes or masks and operate at logic-compatible voltages.
	
	In summary, flash transistors, including FGTs and all variants of CTTs can co-exist with CMOS transistors on the same substrate, in a high-yield, cost-effective manufacturing flow. In this paper, their use will be for realizing logic as opposed to just storage.  The use of flash transistors in logic design was described in~\cite{Abusultan_ICCAD_2016,Abusultan_ICCD_2016,Abusultan_ISVLSI_2016,Abusultan_GLSVLSI_2017}, where the authors demonstrated substantial improvements in power, performance, and area over conventional CMOS standard-cell based design and resilience to aging by reprogramming the $VT$s of the flash transistors when aging-related speed degradation occurred.  The main drawbacks of the approach are (1)~the circuit structures are not compatible with the standard-cell based design flow that is practiced in industry and (2)~they are potentially subject to the similar read/write disturb issues found in memory applications.
	
	\begin{figure*}[ht!]
		\centering
		\includegraphics[width=\textwidth]{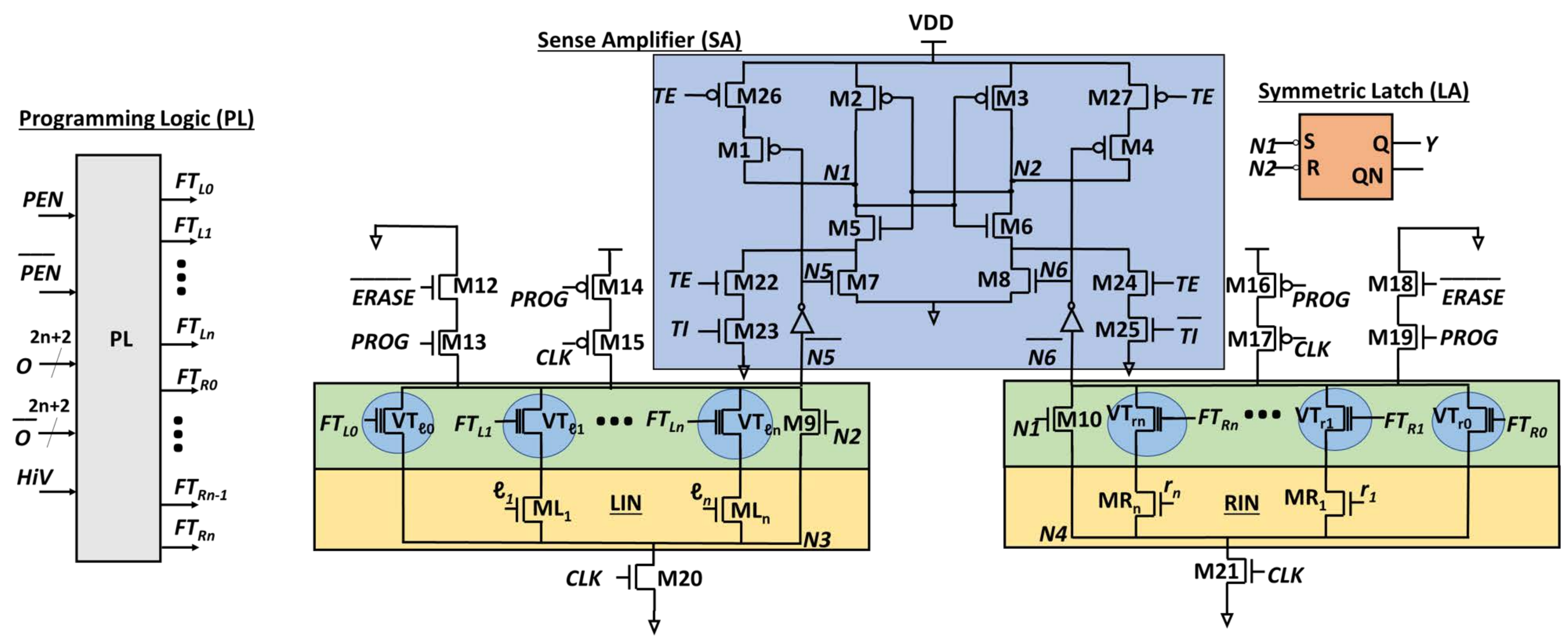}
		\caption{\small FTL Cell Architecture showing LIN, RIN, Sense Amplifier (SA), Latch (LA), and Programming Logic (PL).}
		\label{fig:flash_tlg}
		
		\vspace{-10pt}
	\end{figure*}

	In this paper, we describe how flash transistors can be used to realize threshold logic gates. The flash devices are used as resistors (by varying their threshold voltages) and the resistances serve as a proxy for the weights in a threshold function.  This concept first appeared in~\cite{Bohossian_CPMT_1998}. Their design was a stand-alone, custom-designed analog circuit to realize a 16-input threshold function.  In contrast, threshold gates described in this paper are designed as standard cells, and automatically incorporated in large-scale ASIC design using commercial tools. 
	
	\section{Flash Threshold Logic (FTL) Cell}
	\label{sec:FTLArchitecture}
	
	Figure~\ref{fig:flash_tlg} shows the architecture of the FTL cell.  It has five main components: (i)~the left input network (LIN), (ii)~the right input network (RIN), (iii)~a sense amplifier (SA), (iv)~ an output latch (LA), and (v)~a flash transistor programming logic (PL).  The LIN and RIN consist of two sets of inputs $(\ell_1, \cdots, \ell_n)$ and $(r_1, \cdots, r_n)$, respectively, with each input in series with a flash transistor. In our implementation, $\ell_i=r_i=x_i$ for all $i$. The conductance of the LIN and RIN is determined by the state of the inputs and the threshold voltages of the flash transistors. The assignment of signals to the LIN and RIN is done to ensure sufficient difference in their conductance across all minterms.
	
	There are two differential signals $N1$ and $N2$ in an FTL cell, which serve as inputs to an SR latch.  When $[N1, N2] = [0,1]$ ($[1,0]$), the latch is set (reset) and the output $Y = 1 (0)$.  The magnitudes of the two sides of the inequality in the definition of a threshold function (see Equation~\ref{eq:ThresholdDefinition}) are mapped to the \textit{conductance} $G_L$ of the LIN and $G_R$ of the RIN. Ideally, the mapping is such that $[N1, N2] = [0,1] \Leftrightarrow G_L > G_R$ and $[N1, N2] = [1,0] \Leftrightarrow G_L < G_R$.  As stated earlier, the flash transistor threshold voltages serve as a proxy to the weights of the threshold function~--~the higher the weight, the lower will be the threshold voltage. For a given threshold function, this non-linear monotonic relationship is \textit{learnt} using a modified perceptron learning algorithm described in Section~\ref{sec:WeightsToVTs}.
	
	\begin{figure*}[ht!]
		\centering
		\includegraphics[width=0.9\textwidth]{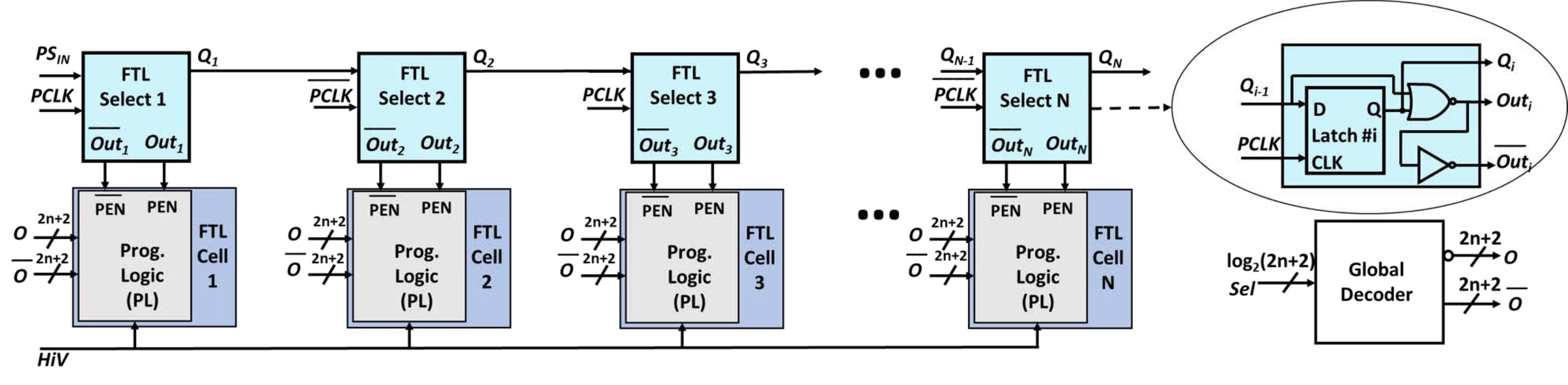}
		\caption{\small Programming Scan Chain for FTL cells in an ASIC.}
		\label{fig:flash_prog_arch}
		\vspace{-10pt}
	\end{figure*}
	
	The FTL cell has four modes: \textit{regular}, \textit{erase}, \textit{programming} and \textit{scan-testing} mode. The $VT$ values of the flash transistors are set in the programming mode and erased in the erase mode. The logic operation of an FTL cell  takes place in regular mode. 
	
	\noindent \textbf{FTL Regular Mode:} ($\textit{PROG} = \textit{ERASE} = 0$, $\textit{TE}=0, \textit{HiV}=0$). Assume that the $VT$s of the flash transistors have been set to appropriate values corresponding to the weights of the threshold function, and their gates are being driven to 1 by setting $FT_i$ to VDD. When $CLK = 0$, the circuit is reset.  In this phase, the nodes $\overline{N5}$ and $\overline{N6}$ of LIN and RIN are connected to the supply, $N5 = N6 = 0$, and $N1 = N2 = 1$. Therefore, the output $Y$ remains unchanged. 
	
	Assume now that an on-set minterm is applied to the inputs in the LIN and RIN.  With properly assigned $VT$ values to the flash transistors, suppose that $G_L > G_R$ for the given minterm.  When  $CLK: 0 \rightarrow 1$, both the LIN and RIN will conduct, and $N5$ and $N6$ will both transition from $0 \rightarrow 1$. Assuming $G_L > G_R$, $N5$ rises faster than $N6$, and hence $N5$ will make $M7$ active before $N6$ makes $M8$ active. This will start to discharge $N1$ before $N2$. When $N1$ falls below the threshold voltage of $M6$, it will stop further discharge of $N2$, and turn on $M3$, resulting in $N2: 0 \rightarrow 1$.  Finally, [N1,N2] = [0,1] sets the SR latch, resulting in $Y = 1$.  For an off-set minterm, $G_L < G_R$, and $[N1, N2]  = [1,0]$ resulting in $Y = 0$. 
	
	\noindent \textbf{FTL Program, Erase and Scan-testing mode}: Figure~\ref{fig:flash_tlg} shows a circuit block labeled PL (programming logic) that generates the signals to select and program the FTL cells at the chip-level. Details of the programming architecture and protocol are given in Section~\ref{sec:ftl_asic_integration}.  During flash-programming of a single FTL, the PL redirects $HiV$ to $FT\textsubscript{i}$, to program the $i\textsuperscript{th}$ flash transistor.
	
	\noindent \textbf{FTL Programming Mode}: (\textit{ERASE}=0, \textit{PROG}=1, \textit{CLK}=0, \textit{HiV}=20V, \textit{TE}=0). The ERASE and PROG signals turn on M12 and M13 and turn off M14. In this state, the source of the flash transistor is floating while the drain and bulk are connected to the ground. Activating the appropriate signals in the PL unit causes high voltage pulses to be applied to the $HiV$ line and  the gate of the flash transistor to set the desired threshold voltage ($VT$).
	
	\noindent \textbf{FTL Erase Mode}: (\textit{ERASE}=1, \textit{PROG}=1, \textit{CLK}=0, \textit{HiV}=-20V, \textit{TE}=0). M12 is turned off by the ERASE signal. Both the source and drain of the flash transistors are floating in this state, while the bulk is connected to the ground. Using the PL unit, the gate terminals of all the flash transistors in the FTL are connected to $HiV$. A negative high voltage pulse at $HiV$ in this state will tunnel the charge from the floating gate, thereby erasing the flash transistor. 
	
	\noindent \textbf{FTL Scan-testing Mode}: (\textit{ERASE}=0, \textit{PROG}=0, \textit{CLK}=0, \textit{HiV}=0, \textit{TE}=1). The scan-testing mechanism in the FTL cells is implemented in the same way as described in \cite{Kulkarni_TVLSI_2016}.  It uses the \textit{test enable} (TE) and \textit{test input} ($TI$ and $\bar{TI}$) signals. In this mode, $TE$ acts as the clock with the main clock $CLK = 0$.  Hence  the scan-testing chains for the D flip-flops and FTL cells are kept separate.  The procedure to inject data into the scan-testing chain of FTL cells is straightforward.  On each $TE$ cycle, the bit to be scanned in is applied to $TI$. Then $TE: 0 \rightarrow 1$, which causes the either $N1$ or $N2$ to discharge resulting in the output latch being set or reset. This process is repeated to load all FTLs with the data in a test vector. Transistors $M26$ and $M27$ block potential DC paths from VDD to VSS during testing.
	
	Note that the problem of read and write disturb~\cite{Bez2003,rdisturb} found in NAND flash memories does not exist with an FTL cell because there is only one flash transistor in each stack in the input network. Also the problem of \textit{write endurance}~\cite{Boboila_FAST_2010} in flash memories, which refers to a limit on the number of writes (from 10K - 100K cycles), is not an issue with FTL cells, because an FTL cell would be programmed or erased only a handful of times over the life of the design. 
	
	\section{Architecture for programming FTL cells}
	\label{sec:ftl_asic_integration}
	
	In this section, we describe the programming architecture used for FTL cells embedded in an ASIC.  This architecture individually addresses the flash transistors of the FTL cells and then redirects HiV pulses to them. Although this architecture is a part of the ASIC, its use ends once the FTL cells are programmed. Therefore, its design must meet its functional requirements without severely impacting the overall area and wirelength of the ASIC.  This is achieved by a scan-chain programming architecture.  
	
	Figure~\ref{fig:flash_prog_arch} shows the structure of the programming scan chain. Each stage of this chain consists of an FTL cell with its programming logic and a select cell that identifies the FTL cell to be programmed. Suppose that the FTL cells realize all threshold functions of $n$ or fewer variables.\footnote{In the experimental results, $n = 5$.}  Then each such cell has $2n+2$ flash transistors.  Suppose further that there are $N$ FTL cells. Initially, all $Q_i$s are set to 1.  Then cell $i$ is selected by making $Q_i$ = $Q_{i-1} = 0$, while all other $Q$s remain at 1. Thus, clocking in the appropriate sequence using PCLK selects cell $i$. Since each FTL cell has $2n+2$ flash cells, a global decoder with $\log(2n+2)$ inputs and $2n+2$ outputs is used. These outputs of the decoder activate the appropriate path for the HiV pulses to the inputs of the flash transistors of the selected FTL cell. 
	
	The programming architecture involves the use of high voltage nets. In the physical layout, the high voltage wires are bundled, and wire-shielding \cite{mehri_thorough_2015} is used to avoid any cross-talk due to high voltage signals to the other low voltage lines and transistors. Programming is done with a dedicated scan chain, and all the associated high voltage nets are systematically bundled and shielded. This results in reducing the total wirelength of those nets. Furthermore, since it is a linear iterative array, it easily scales to accommodate any number of cells. 
	
	Assuming that FTL cells use floating gate transistors, the program and erase modes require +20V and -20V (HiV)  pulses to be applied to their inputs (see Section \ref{subsec:background:flash}). Note that other flash technologies such as SONOS~\cite{nii_cost-effective_2020}, MONOS\cite{tsuda_first_2016} and High-K Metal Gate (HKMG) \cite{khan_turning_2019} require similar infrastructure for programming and erasure, but may differ in the voltage levels of the pulses.
	
	\section{Computing the relationship between the weights and the $V_T$ values for an FTL cell}
	\label{sec:WeightsToVTs}
	
	\subsection{Overview}
	
	Let  $\bm{VT}_\ell(f) = (VT_{\ell_0}(f),\cdots, VT_{\ell_n}(f))$, and $\bm{VT}_{r}(f) = (VT_{r_0}(f), \cdots, VT_{r_n}(f))$  denote the threshold voltages of the flash transistors in the LIN and RIN of an FTL, respectively (see Figure~\ref{fig:flash_tlg}). Further, let $\bm{VT}(f) = (\bm{VT}_\ell(f), \bm{VT}_r(f))$. In this section, we present an algorithm to determine these voltages for an FTL to realize a given threshold function $f$ having a weight vector $[\bm{W}, T]$.
	
	To configure an FTL, the method described herein determines $\bm{VT}(f)$ for each $f$ a priori, using models that include circuit parasitics and global and local process variations in the device and circuit parameters.   This is to ensure that an overwhelming majority ($\gg 99\%$) of the FTL instances on a chip can be programmed by attempting at most a few pre-computed values of $\bm{VT}(f)$. The remaining small fraction of FTLs for which a feasible, model-based  $\bm{VT}(f)$ could not be found, can be programmed directly on the chip. 
	
	Let $G_L(x|\bm{VT}(f))$ and $G_R(x|\bm{VT}(f))$ for $x \in B^n$, denote the conductance of the LIN and RIN as functions of a minterm $x$ of $f$ and the flash transistor threshold voltages. Henceforth, for clarity, we refer to  $G_L(x|\bm{VT}(f))$ and $G_R(x|\bm{VT}(f))$ as $G_L$ and $G_R$ respectively. 
	
\begin{figure}[ht]
		\centerline{\includegraphics[width=\columnwidth]{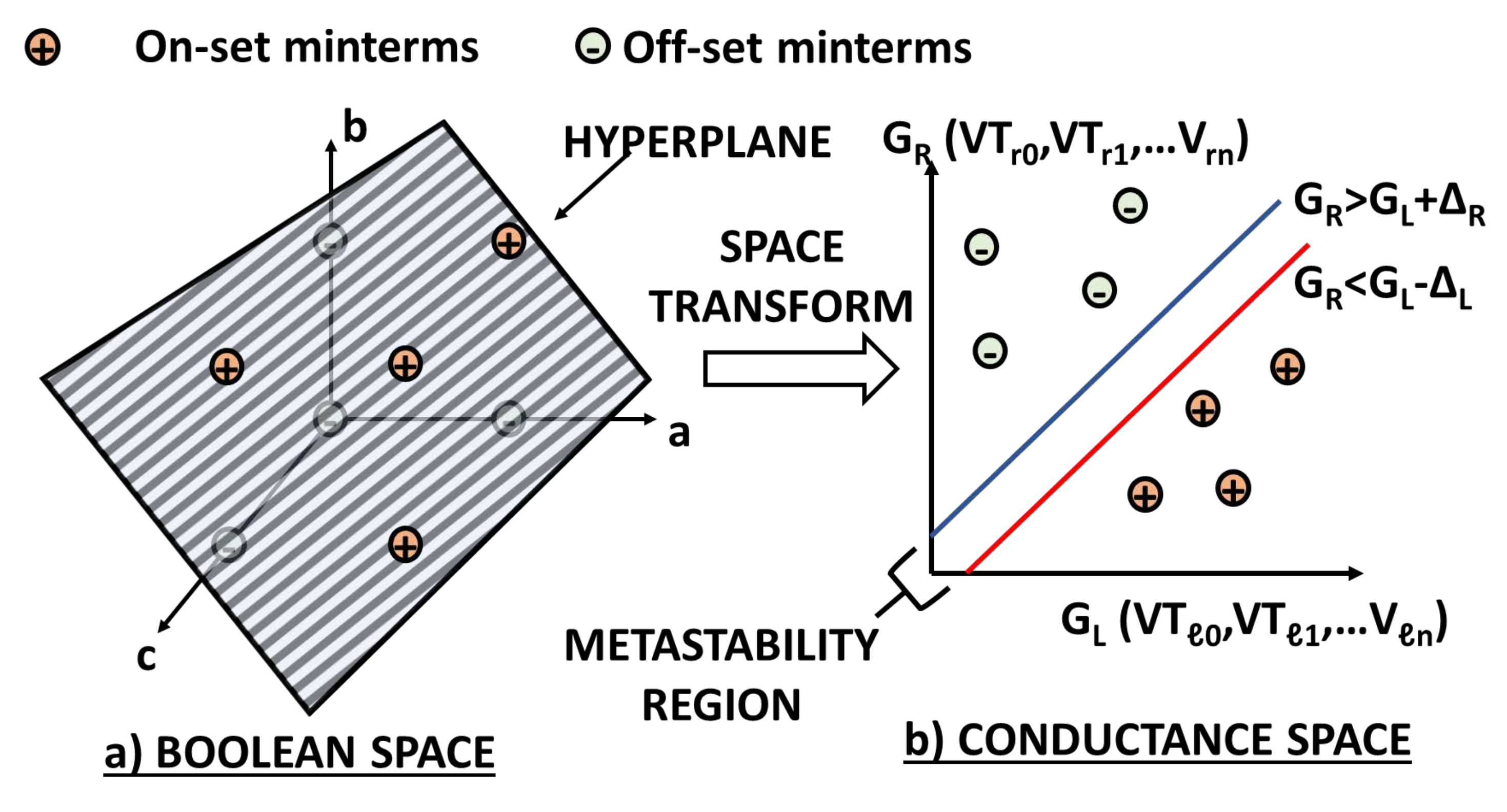}}
		\caption{\small Transformation from Boolean space to conductance space.}
		\label{fig:space_transformation}
	\end{figure}

	The problem is to find a $\bm{VT}(f)$ that determines a mapping between the Boolean space $B^n$ and the conductance space $(G_L, G_R)$ such that, in the \textit{ideal} case, 
	\begin{equation}
		\label{eq:conductance_space_def}
		\begin{split}
			G_R < G_L, \text{~if~} f(x) = 1 \\
			G_R > G_L, \text{~if~} f(x) = 0.
		\end{split}
	\end{equation}
	This mapping, depicted in Figure~\ref{fig:space_transformation}, is one-to-many, since there can be many (an uncountable number) \textit{feasible} values of $\bm{VT}(f)$ for a given $f$ with a weight vector $[\bm{W}, T]$.  
	
	In practice, to avoid variations due to parasitics which could make the circuit behavior erroneous, we require finding a subset of the feasible set where 
	\begin{equation}
		\label{eq:conductance_space}
		\begin{split}
			G_R < G_L - \Delta_L, \text{~if~} f(x) = 1 \\
			G_R > G_L + \Delta_R, \text{~if~} f(x) = 0
		\end{split}
	\end{equation}
	for some some sufficiently large $\Delta_L \in  (0, G_L)$ and $\Delta_R \in (0, G_R)$. Note that $\Delta_L$ and $\Delta_R$ are related due to the constraints imposed by the truth table. 
	
	Our approach to find $\bm{VT}(f)$ consists of three steps which are implemented by Algorithms~\ref{alg:mPLA0}, $\text{mPLA}^+$, and \ref{alg:mPLA++}. These are described in the following sections. 
	
	\subsection{Algorithm $\text{mPLA}^{0}$}
	\label{subsec:mPLA}
	
	Algorithm~\ref{alg:mPLA0} is a modified version of the classical perceptron learning algorithm (PLA)~\cite{rosenblatt_PR_1958} that works in concert with HSPICE (for verifying the truth table of $f$) to search through the space of values of $\bm{VT}(f)$ until each minterm (a  point in the $(G_L, G_R)$ space) of $f$ is correctly classified.  It does this by iteratively adjusting the threshold voltages of flash transistors such that points in the conductance space that correspond to the on-set and off-set minterms satisfy the constraints in Equation~\ref{eq:conductance_space} (see Figure \ref{fig:space_transformation}). It calls HSPICE (line 3 of Algorithm \ref{alg:mPLA0}) to determine whether any point falls above or below the lines corresponding to these inequalities.   Since a layout extracted FTL circuit is being used,  the circuit parasitics are accounted for in the HSPICE simulation.  Consequently, Algorithm~\ref{alg:mPLA0} implicitly finds values of $\Delta_L$ and $\Delta_R$.  
	
	\begin{varalgorithm}{$\text{mPLA}^{0}$}
		\caption{\small Modified Perceptron Learning Algorithm}
		\label{alg:mPLA0}
		\small
		\begin{algorithmic}[1]
			\renewcommand{\algorithmicrequire}{\textbf{Input: Truth table TT of threshold function $f$}}
			\renewcommand{\algorithmicensure}{\textbf{Output: $\bm{VT}_0$ to program FTL cells with $f$}}
			\REQUIRE
			\ENSURE
			\STATE Initialize $\bm{VT}_0$
			\FOR {$k$ = $0$ to $kmax-1$}
			\STATE OT = SPICE($\bm{VT}_0$) // Truth table from simulation
			\IF {TT and OT disagree on some minterm m}
			\IF {TT(m)==1}
			\STATE {Update $\bm{VT}_0$: decrement (increment) the $VT$ of every active transistor in LIN (RIN) that is '1' in m by $\delta$}
			\ELSE
			\STATE {Update $\bm{VT}_0$: increment (decrement) the $V_T$ of every active transistor in LIN (RIN) that is '1' in m by $\delta$}
			\ENDIF
			\ELSE 
			\STATE Break
			\ENDIF
			\ENDFOR
		\end{algorithmic}
	\end{varalgorithm}
	
	Given the truth table ($TT$) of $f$, $\text{mPLA}^{0}$ applies all the minterms of $f$ to the FTL cell and records the HSPICE response in $OT$ (output table). If $TT(m)\neq OT(m)$, for some minterm $m$, then the constraint in Equation~\ref{eq:conductance_space} was not satisfied. In such a case, $\text{mPLA}^{0}$ adjusts the values of $\bm{VT}^0(f)$ (Algorithm \ref{alg:mPLA0} line 4-9) associated with the active input transistors within the interval $[\delta, V_{DD} - \delta]$, by a minimum increment $\delta$, according to Equation~\ref{eq:vt_update}. Here, $m_\ell$ ($m_r$) is a binary vector that identifies the active input transistors in the LIN (RIN).
	
	\begin{equation}
		\label{eq:vt_update}
		\small
		\begin{split}
			\textbf{VT}_{\ell}^{k+1} =
			\begin{cases}
				\textbf{VT}_{\ell}^{k} - \delta m_\ell & \text{ if } m \cdot W \ge T\\
				\textbf{VT}_{\ell}^{k} + \delta m_\ell & \text{ if } m \cdot W < T
			\end{cases}\\
			\textbf{VT}_{r}^{k+1} =
			\begin{cases}
				\textbf{VT}_{r}^{k} + \delta m_r & \text{ if } m \cdot W \ge T\\
				\textbf{VT}_{r}^{k} - \delta m_r & \text{ if } m \cdot W < T
			\end{cases}
		\end{split}
	\end{equation}
	
	The term $\delta m_l$ (or $\delta m_r$) is a vector which has a value $\delta$ at all locations in LIN (RIN) which are 1 for a minterm $m$, and zero elsewhere. For instance, consider the threshold function $b+c\ge a+T$. Let $m=110$ be an on-set minterm that was incorrectly evaluated. If $TT(m)\neq OT(m)$ then $G_R > G_L - \Delta_L$.  Therefore $G_L$ needs to be increased (threshold voltages corresponding to the flash transistors of $b$ and $c$ will be decreased) and $G_R$ needs to be decreased (threshold voltages corresponding to the flash transistors of $a$ and $T$ will be increased) for minterm $m$. Consequently, the threshold voltages of all the flash transistors associated with the active input transistors should be decreased (increased) by $\delta$ in the LIN (RIN). A similar change is made if $m$ is an off-set minterm. This is what is expressed in Equation~(\ref{eq:vt_update}).  $\bm{VT}^{0}(f)$ is the value returned by Algorithm~\ref{alg:mPLA0}.

	If a given set of points in $B^n$ is linearly separable (i.e. a threshold function), then the PLA algorithm will terminate in a finite number iterations~\cite{Siu_1995,McCulloch:1988}.  Similarly, given a threshold function $f$, a sufficiently small $\delta$ and an FTL instance for which there exists a feasible $\bm{VT}(f)$,  Algorithm~\ref{alg:mPLA0} will terminate in a finite number steps  (see ~\cite{Siu_1995} for proof of termination). For an $n$-input threshold function, the upper bound on the number of iterations of the PLA given in~\cite{Siu_1995} becomes $kmax = 2(n+1)||\bm{VT}^{0}(f)||^2/\delta^2$.  For instance, with $n = 5$ and $\delta = .02V$, $kmax = 3000||\bm{VT}^{0}(f)||^2$. 
	
	\subsection{Algorithm $\text{mPLA}^{+}$: Improving Noise Tolerance}
	\label{robustness_training}
	
	Algorithm~\ref{alg:mPLA0}  does not consider the relative location of the points with respect to the metastability region defined by the lines $G_R=G_L-\Delta_L$ and $G_R=G_L+\Delta_R$ (see Figure~\ref{fig:space_transformation}b). Even though minterms are classified correctly, they can be arbitrarily close to the metastability region. The further a minterm is from this region, the easier (and faster and more robust) it will be for the sense amplifier to detect the difference between $N5$ and $N6$, and discharge the appropriate side ($N1$ or $N2$) first.  Thus, maximizing $\Delta_L$ and $\Delta_R$ within the feasible set will maximize its noise tolerance.  
	
	Algorithm $\text{mPLA}^{+}$ repeatedly calls $\text{mPLA}^0$ to maximize $\Delta_L$ and $\Delta_R$.  It does this by introducing a \textit{hypothetical} capacitance $C_1$ on node $\overline{N5}$ (which is introduced in HSPICE) when classifying an on-set minterm, and determining the maximum value of $C_1$ for which Algorithm~\ref{alg:mPLA0} converges. This modification \textit{handicaps} node $\overline{N5}$ and directs the algorithm to find a solution, that will result in an increased gap between $G_L$ and $G_R$. Similarly, we add a capacitance $C_0$ on node $\overline{N6}$, when classifying an off-set minterm. Since the values of $\Delta_L$ and $\Delta_R$ are linearly proportional to $C_1$ and $C_0$ respectively, the separation between the lines $G_R = G_L-\Delta_L$ and $G_R = G_L+\Delta_R$ increases, which in turn forces the training algorithm to produce a threshold voltage assignment $\bm{VT}^+(f)$ in a more robust (and also faster) FTL cell. Note that $C_1$ and $C_0$ are only used during HSPICE simulations, and are not part of the actual FTL cell.
	
\begin{figure}[ht]
		\centerline{\includegraphics[width=\columnwidth]{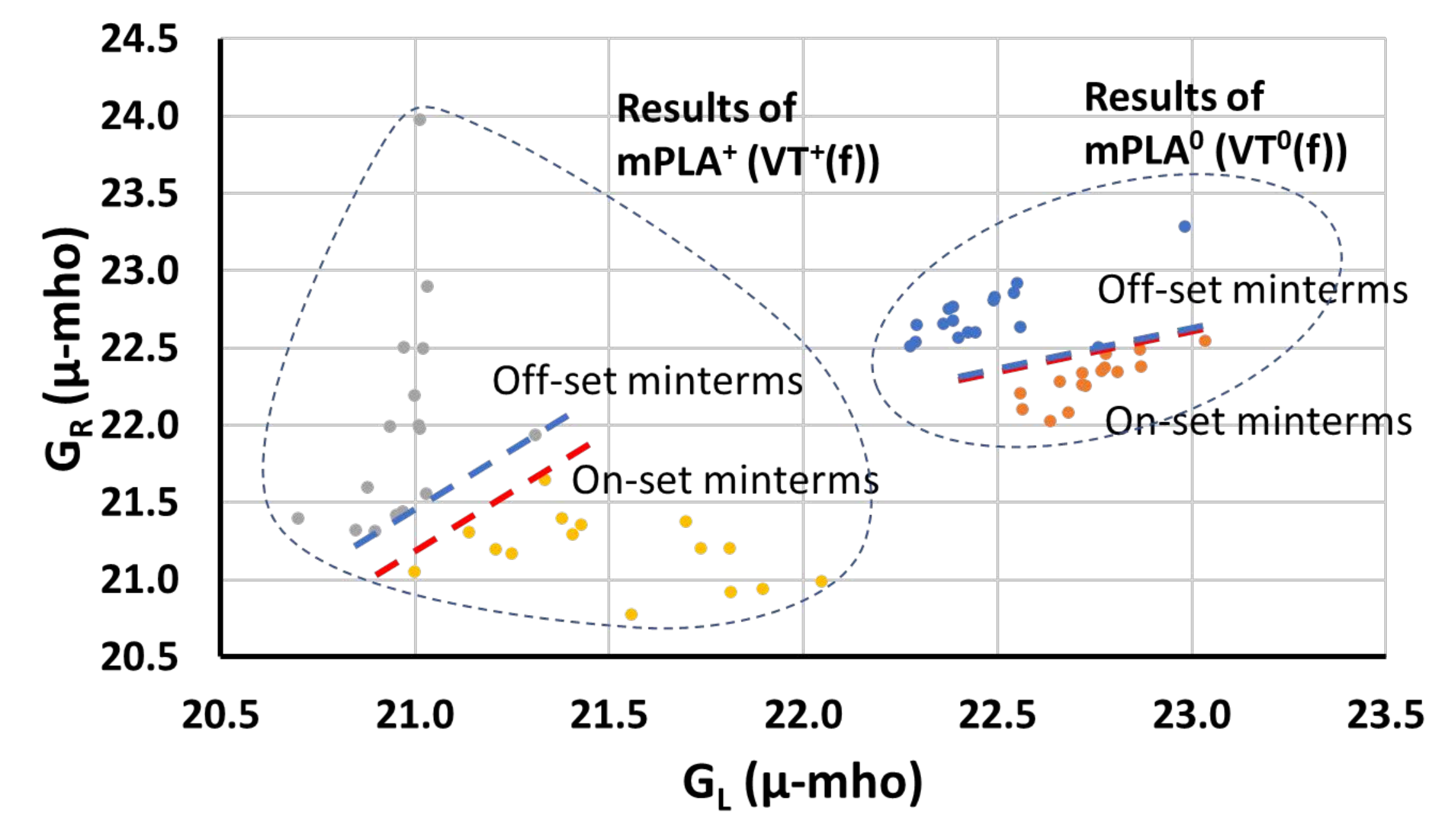}}
		\vspace{-10pt}
		\caption{\small Conductance $G_L$ and $G_R$ of FTL cell programmed for $f = [4, 1, 1, 1, 1; 5]$  using $\text{mPLA}^{0}$ and $\text{mPLA}^{+}$ ($[TT, 0.9V, 25^{\circ}C]$).}
		\label{fig:conductance_space_real}
	\end{figure}

	Figure~\ref{fig:conductance_space_real} shows the results of running Algorithms~\ref{alg:mPLA0} and  $\text{mPLA}^{+}$ on a test function (\cite{MurogaBook71}) $f_{115}(a,b,c,d,e): (\bm{W}, T) = [4, 1, 1, 1, 1; 5] =  a(b+c+d+e)$. It is plot of the minterms in the conductivity space that was obtained by using HSPICE after programming the FTL using $\bm{VT}^{0}(f)$ and $\bm{VT}^{+}(f)$. The largest values of $C_1$ and $C_0$ for which a feasible solution was obtained was $0.1fF$.  The plot shows that training with the hypothetical capacitance values separates the two closest on-set and off-set minterms in the conductivity space by more than five times.  Furthermore, the delay of an FTL programmed with $\bm{VT}^{+}(f)$ will be smaller than the one that is programmed with $\bm{VT}^{0}(f)$.

	\subsection{Algorithm~\ref{alg:mPLA++}: Optimizing Yield}
	\label{subsec:YieldOptimization}
	
	The threshold voltages $\bm{VT}^+(f)$ computed by the $\text{mPLA}^+$ are aimed at achieving maximal separation between the on-set and off-set minterms based on model-based estimates of parasitics. This has the twin advantages of increasing the noise margin and reducing the delay. Despite this, inevitable manufacturing variations can still result in reducing the difference between $G_L$ and $G_R$ associated with $\bm{VT}^{+}(f)$ of the two closest minterms, which may result in the incorrect evaluation of the intended threshold function.  In this section we present a \textit{predictive} technique to \textit{pre-compute} a small \textbf{set} of $\bm{VT}s(f)$ for each threshold function $f$ which would cover a very high percentage of manufactured variations.  
	
	Among the $N$ manufactured FTL cells  programmed to realize function $f$ using $\bm{VT}^{+}(f)$, suppose that $N_e$ were erroneous and let $\{\text{FTL}_1(f), \text{FTL}_2(f), \cdots, \text{FTL}_{N_e}(f)\}$ be the erroneous instances.  The problem is to find individual threshold voltage assignments for each of these $N_e$ instances so that each will correctly realize $f$. Our approach is motivated by two observations. 
	
	First, each erroneous function in  $\{f^{e}_{i}, 1 \leq i \le N_e\}$ is itself a threshold function.   This is simply due to the fact that by construction, an FTL cell only computes threshold functions (see Figure~\ref{fig:FTLSchematic}). Second, our experiments show that a large number of different instances of an FTL cell, which are reprogrammed with $\bm{VT}^{+}(f)$ and are to realize the same function $f$, realize the same erroneous function $f^{e}$.   This suggests that all the erroneous FTL cell instances can be grouped into a few equivalence classes, called \textit{error-types}, with two FTLs belonging to the same error-type if they realize the same erroneous function.  
	
	Given a threshold function $f$, Algorithm~\ref{alg:mPLA++} first generates a \textit{set} of $N_{MC}$ Monte Carlo (MC) instances of an FTL cell and identifies the $N_e$ erroneous instances (i.e. those when programmed with $\bm{VT}^{+}(f)$ do not realize $f$).  The $N_e$ erroneous instances are grouped into $M_f$ error-types. Let $f^{e}_{i}, 1 \leq i \leq M_f$, denote the logic functions of the \textit{distinct} error-types observed in a sample of $N$ FTLs.  Algorithm~\ref{alg:mPLA++} selects one MC instance from each error-type class and computes one $\bm{VT}^{+}(f)$ assignment for that instance using $\text{mPLA}^{+}$. It returns a set of threshold assignments, 
	\begin{equation}
		\bm{VT}^{++}(f) = \{\bm{VT}^{+}(f_1^{e}), \bm{VT}^{+}(f_2^{e}), \cdots, \bm{VT}^{+}(f_{M_f}^{e})\}, 
	\end{equation}
	one for each error-type for each function $f$. 
	
	\begin{varalgorithm}{$\text{mPLA}^{++}$}
		\caption{\small Modified Perceptron Learning Algorithm accounting for process variations}
		\label{alg:mPLA++}
		\small
		\begin{algorithmic}[1]
			\renewcommand{\algorithmicrequire}{\textbf{Input: TT of $f$, $N_{MC}$}}
			\renewcommand{\algorithmicensure}{\textbf{Output: $\bm{VT}^{++}(f)$ to program FTL cells with $f$}}
			\REQUIRE
			\ENSURE
			
			\vspace*{0.5em}
			\STATE {Execute $\text{mPLA}^+$ to compute $\bm{VT}^+(f)$}
			\vspace*{0.5em}
			
			\STATE {Using MC sampling of the parameter space, generate $N_{MC}(f)$ instances of an FTL cell, and program them with $\bm{VT}^+(f)$.}
			\vspace*{0.5em}
			
			\STATE {Among the set of $N_{MC}$ instances, let $N_e$ be the number of instances, which when programmed with $\bm{VT}^+(f)$, realize a function other than $f$, and among these $N_e$, let $M_f$ be the number of erroneous functions that are distinct.}
			\vspace*{0.5em}
			
			\STATE { Execute the $\text{mPLA}^+$  on one MC instance from each of the $M_f$ erroneous functions to obtain\\
				\begin{center} 
					$\{\bm{VT}^{++}(f)\} = \{\bm{VT}^+(f^e_1), \cdots, \bm{VT}^+(f^e_{M_f}).\}$
				\end{center}
			}
			\vspace*{0.5em}
		\end{algorithmic}
	\end{varalgorithm}
	
	Results presented in Section~\ref{sec:ExpResults} show that using the $\bm{VT}^{+}(f^e_i)$ computed for one FTL instance from $i^{th}$ error-type ($1 \leq i \leq M_f$) resulted in {\em{all}} the instances of the same error-type correctly realizing $f$. This works because instances that have the same error-type share similar parasitic variations. Thus, all instances of our sample of FTL cells were correctly programmed using one distinct $\bm{VT}^{+}(f^e_i)$ for each error-type. 
	
	There is no guarantee that the set of erroneous functions found in a sample set $N_{MC}$ will capture all manufacturing outcomes.  This means that there may be some manufactured FTLs that could not be correctly programmed using any of the threshold voltage vectors computed by Algorithm~\ref{alg:mPLA++}.  For these remaining FTLs, our approach is to 
	apply Algorithm~\ref{alg:mPLA0} directly on the chip. In each iteration of $\text{mPLA}^0$, the step that adjusts the threshold voltages of flash transistors is replaced by the application of an appropriate number of positive or negative pulses to the FTL cell on the chip using the programming scan chain. This capability of correcting the function of a cell after fabrication to increase yield is a signature attribute of the proposed design methodology.

	\section{Experimental Results}
	\label{sec:ExpResults}
	
	\subsection{Experiment Setup}
	
	An FTL cell with $n = 5$ (see Figure~\ref{fig:flash_tlg} in Section~\ref{sec:FTLArchitecture}) was designed and a complete layout (including the programming devices) was created using the TSMC 40nm LP library. It was laid out as a double height cell requiring 24 tracks. The flash transistor models were obtained from \cite{Abusultan_ICCD_2016} and were suitably modified to reflect the characteristics and variations of the TSMC 40nm LP library. The design rules for the flash transistors were obtained from ITRS. The standard cell area of the FTL was15.6 $\mu m^2$. 
	
	There are a total of 117 distinct positive-form threshold functions of five or fewer variables.  A numbered list of these is given in~\cite{MurogaBook71} and can also be accessed at~\cite{Threshold5List}.  The one cell that was designed was copied 117 times, and each was trained to realize one of the 117 threshold functions. In this section, we use the same numbering scheme as in~\cite{MurogaBook71} to identify the functions. The FTL cell trained to implement the threshold function numbered $n$ in~\cite{MurogaBook71} will be referred to as $FTL_{n}$, and the corresponding CMOS implementation will be denoted as $CMOS_{n}$.  The threshold function itself will be denoted as $f_{n}$.  \textbf{Note:} In all the bar charts shown in this section, the numbers on the x-axis identify the threshold function.  Function $f_0$ is a buffer and is omitted because this would correspond to a DFF, which by itself would never replaced by an FTL in an ASIC.  The first function shown is $f_1$, which is a two-input AND.
	
	\subsection{Training Iterations}
	
\begin{figure}[ht]
		\centering
		\includegraphics[width=0.9\columnwidth]{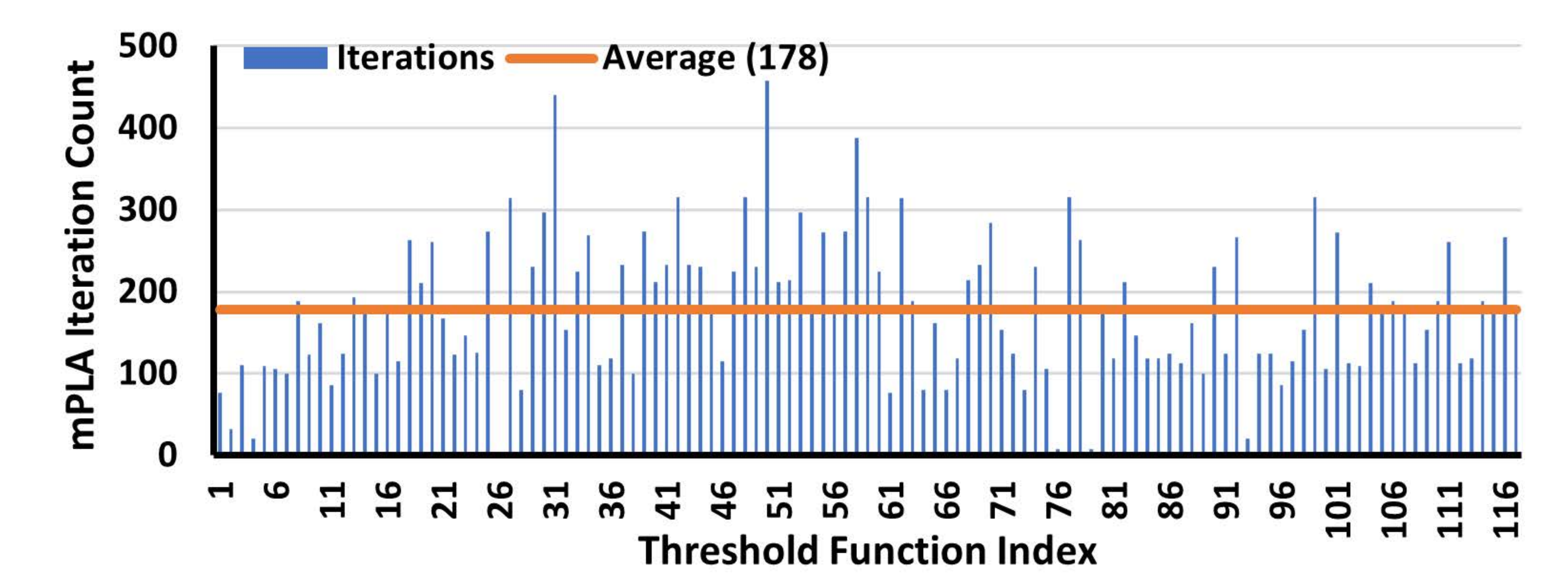}
		\caption{\small Iteration count for $\text{mPLA}^+$ for all 117 functions of 5 or fewer variables.}
		\label{fig:iteration_count}
		\vspace{-10pt}
	\end{figure}

	$\text{mPLA}^+$ was used to train the FTL cell for robustness (see Section~\ref{robustness_training}) for all 117 functions.  Figure~\ref{fig:iteration_count} shows the number of iterations needed for training of each of the 117 functions. The actual number of iterations was about 10X lower than the theoretical upper bound, presented in Section \ref{subsec:mPLA}.
	
	\subsection{Individual Cell Area, Delay and Power Comparison}
	\label{sec:CellPPA}
	
\begin{figure*}[ht]
		\centering 
		\includegraphics[width=\textwidth]{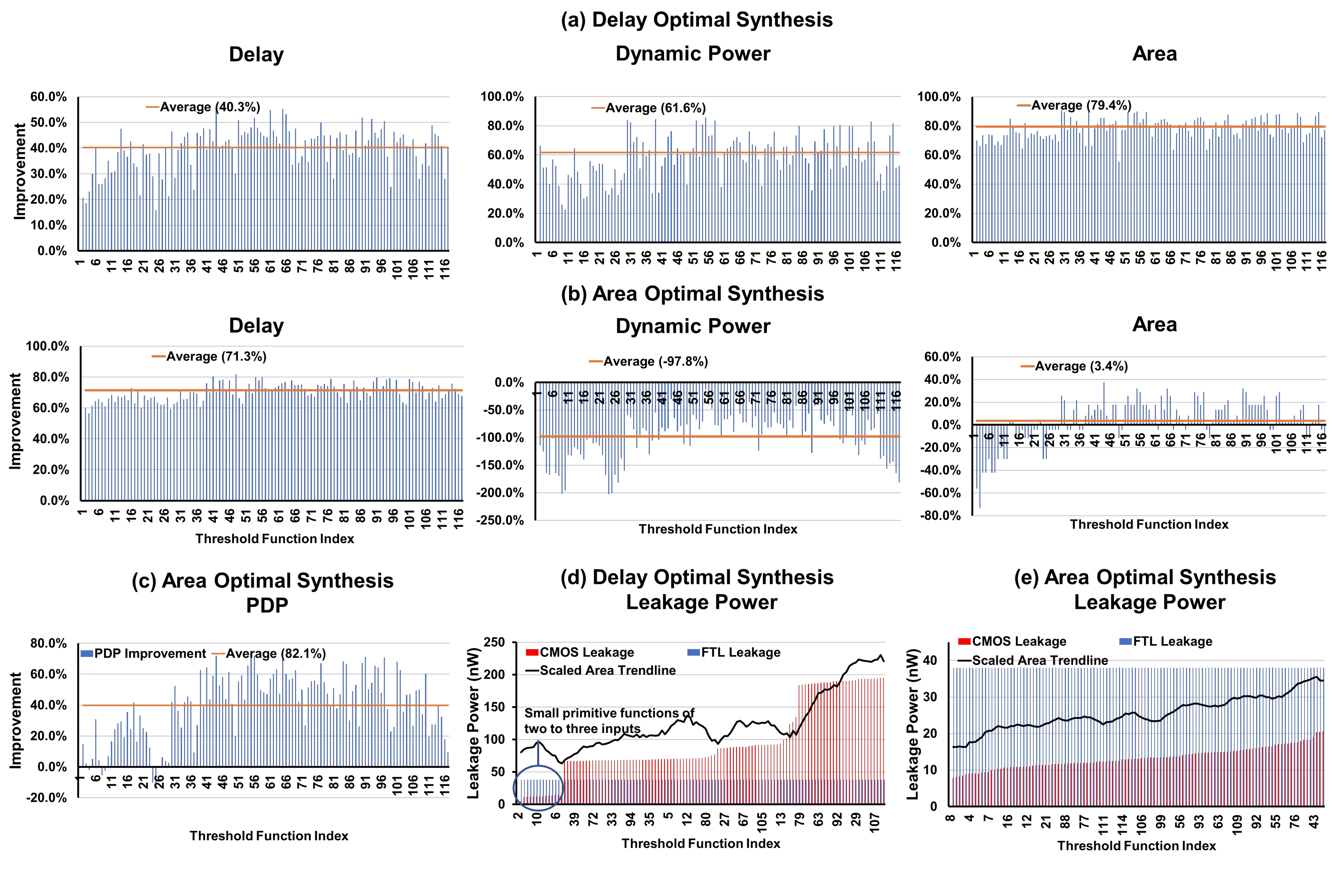}
		\caption{{\small {PPA improvements of FTL over CMOS implementations. Simulations done at 25$^{\circ}$C assuming a 20\% input switching activity.}}}
		\label{fig:combined_improvement} 
	\end{figure*}

	All 117 threshold functions of five or fewer variables were implemented using FTL cells. These functions were also synthesized by Cadence Genus \cite{Cadence} and placed and routed using Cadence Innovus \cite{Cadence}, using the conventional TSMC 40nm LP standard cells. Two sets of experiments were performed to compare the CMOS equivalent designs to the FTL cells: (1)~delay optimal and (2)~area optimal synthesis. The results comparing the total delay (sum of logic delay, setup-time, and clock-to-Q delay), area, and power of these circuits and the corresponding FTL implementations are shown in Figures~\ref{fig:combined_improvement}(a) and \ref{fig:combined_improvement}(b), respectively. 
	
	The results show that FTL cells have the advantage of speed. Optimizing their CMOS equivalents to meet the delay of the corresponding FTL cells forces the synthesis algorithms to use high drive strength cells (larger area) for the combinational logic and larger DFFs.  As the FTL implementations are faster than the fastest CMOS equivalent implementation, delay optimal synthesis results in an across-the-board improvement in all FTL cells in delay, area, and power. 
	
	When synthesizing individual cells for minimum area, FTL cells are still uniformly faster.  However, the synthesis algorithm now uses the smallest combinational cells and DFFs in the CMOS equivalents.  In this case, although the CMOS implementations of simpler functions are much smaller than their FTL equivalents (see Figure~\ref{fig:combined_improvement}(b)), the FLT versions are still smaller for 74 out of 117 functions because the \textit{logic absorbed} by the FTL cell results in greater area savings than the smaller drive strength cells used in the CMOS equivalents.
	
	The dynamic power of every FTL implementation is higher than its CMOS equivalent for area optimal synthesis.  The reasons for this are (1)~an FTL cell resets and then evaluates its function on every clock cycle and (2)~the much smaller switched capacitance of the low-drive strengths of the combinational logic in the CMOS equivalents. Figure~\ref{fig:combined_improvement}(a) shows that FTL cells have a much lower power-delay product (i.e. energy) when their CMOS equivalents are synthesized for minimum delay. Figure~\ref{fig:combined_improvement}(c) shows that this also true for the majority of the CMOS equivalents when they are synthesized for minimum area. Hence, FTL cells are, in general, more energy efficient.
	
	Figures~\ref{fig:combined_improvement}(d) and \ref{fig:combined_improvement}(e) show a comparison of the leakage power of FTL cells and their CMOS equivalents. The leakage of FTL cells is practically independent of the function, and in the case of delay optimal synthesis, it is far lower than every CMOS equivalent circuit. Exactly the opposite is true for the area optimal synthesis due to reduced sizes of the combinational cells and DFFs. In these plots the circuit indices are ordered by increasing area.  The area trend lines show that the leakage increases with area for the CMOS implementations. 
	
	\subsection{Delay Distributions}
	\label{subsec:DelayDistributions}
	
\begin{figure}[h]
		\centering
		\includegraphics[width=0.9\columnwidth]{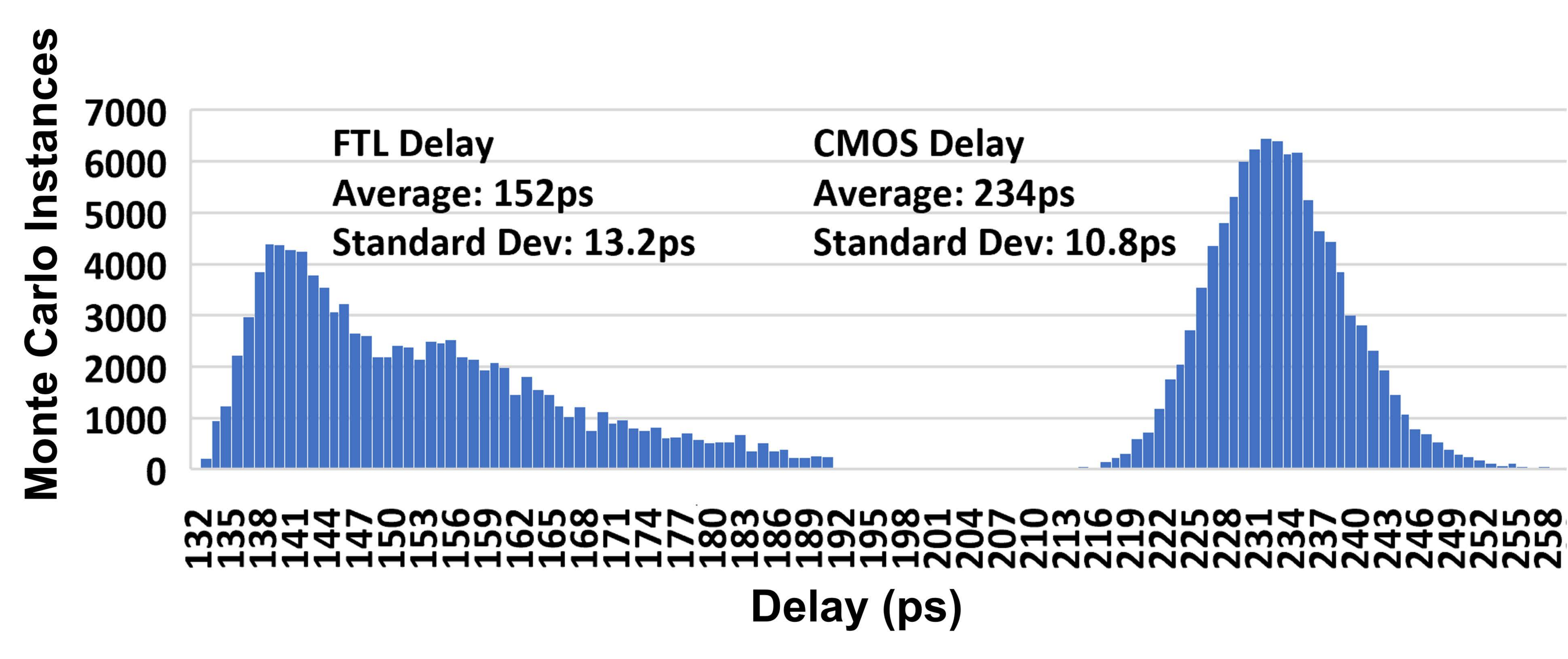}
		\caption{\label{monte_carlo_delay} \small Delay histogram of $FTL_{35}$ and $CMOS_{35}$ with 100K Monte Carlo simulations. $PVT = [TT,0.9 V,25^\circ C]$.}
	\end{figure}

	This experiment compares the distributions of delays of FTL and CMOS implementations. We show the results for the threshold function $f_{35}$ with a weight vector $[\bm{W}; T] = [3, 3, 2, 1, 1; 8]$. The PVT corner setting was $[TT,0.9V,25^\circ C]$.  100K Monte Carlo instances were generated for both $FTL_{35}$ and $CMOS_{35}$.  Each of the 100K FTL instances was verified against the truth table for functional correctness, for both $FTL_{35}$ and $CMOS_{35}$.  Figure~\ref{monte_carlo_delay} shows the histogram of delays for both circuits. These demonstrate the delay advantage of the FTL cell over its CMOS equivalent, even in the presence of process variations.  The difference in standard deviation between the two is insignificant. Note that the FTL instances with large delays can be \textit{re-programmed} to reduce the delay further.  This capability is not possible for the CMOS versions. 
	
	\vspace*{-1em}
	
	\subsection{Dynamic Voltage Scaling}
	\label{subsec:VoltageScaling}
	
	Voltage scaling is a common mechanism to trade off performance against power. Table~\ref{voltage_sweep2} shows the results of training $FTL_{35}$ at $0.9V$.  The FTL cell was programmed with the determined set of flash threshold voltages, and then operated over the voltage range $[0.8V, 1.1V]$. To ensure proper operation across all voltages, the gate voltages of the flash transistors were scaled accordingly. This result demonstrates how a single $\bm{VT}^+(f)$ assignment can be used for dynamic voltage scaling. The delay of the $FTL_{35}$ varies by 2.5X (its CMOS equivalent by 2.8X), power varies by 5.9X (CMOS equivalent by 1.9X), and the PDP (energy) varies by 2.3X (CMOS equivalent by 1.43X), as the supply voltage varies over [0.8V, 1.1V]. 
	
	\begin{table}[h]
		\centering
		\small
		\begin{tabular}{|c|c|c|c|c|}
			\hline
			\begin{tabular}[c]{@{}c@{}}Supply\\ Voltage (V)\end{tabular} & \begin{tabular}[c]{@{}c@{}}Flash Gate\\ Voltage (V)\end{tabular} & \begin{tabular}[c]{@{}c@{}} Power\\ (uW) \end{tabular} & \begin{tabular}[c]{@{}c@{}} Delay\\ (ps) \end{tabular} & PDP    \\ \hline
			0.8         & 0.8                                                              & 14.3      & 198.1      & 2837.1 \\ \hline
			0.85        & 0.825                                                            & 20.5      & 157.6      & 3228.7 \\ \hline
			0.9         & 0.85                                                             & 26.1      & 130.2      & 3396.9 \\ \hline
			0.95        & 0.875                                                            & 40.3      & 111.2      & 4482.7 \\ \hline
			1           & 0.9                                                              & 53.1      & 97.0       & 5148.6 \\ \hline
			1.05        & 0.925                                                            & 76.0      & 86.4       & 6562.9 \\ \hline
			1.1         & 0.95                                                             & 85.0      & 78.2       & 6644.0 \\ \hline
		\end{tabular}
		\caption{	\label{voltage_sweep2}\small Delay, total power and power-delay-product (PDP) of $FTL_{35}$, trained at $V_{DD} = 0.9V$, and $C_{0} = C_1 = 0.1fF$.}
		\vspace{-10pt}
	\end{table}

	\subsection{Number of programming pulses}
	
\begin{figure}[ht]
		\centering 
		\includegraphics[width=0.9\columnwidth]{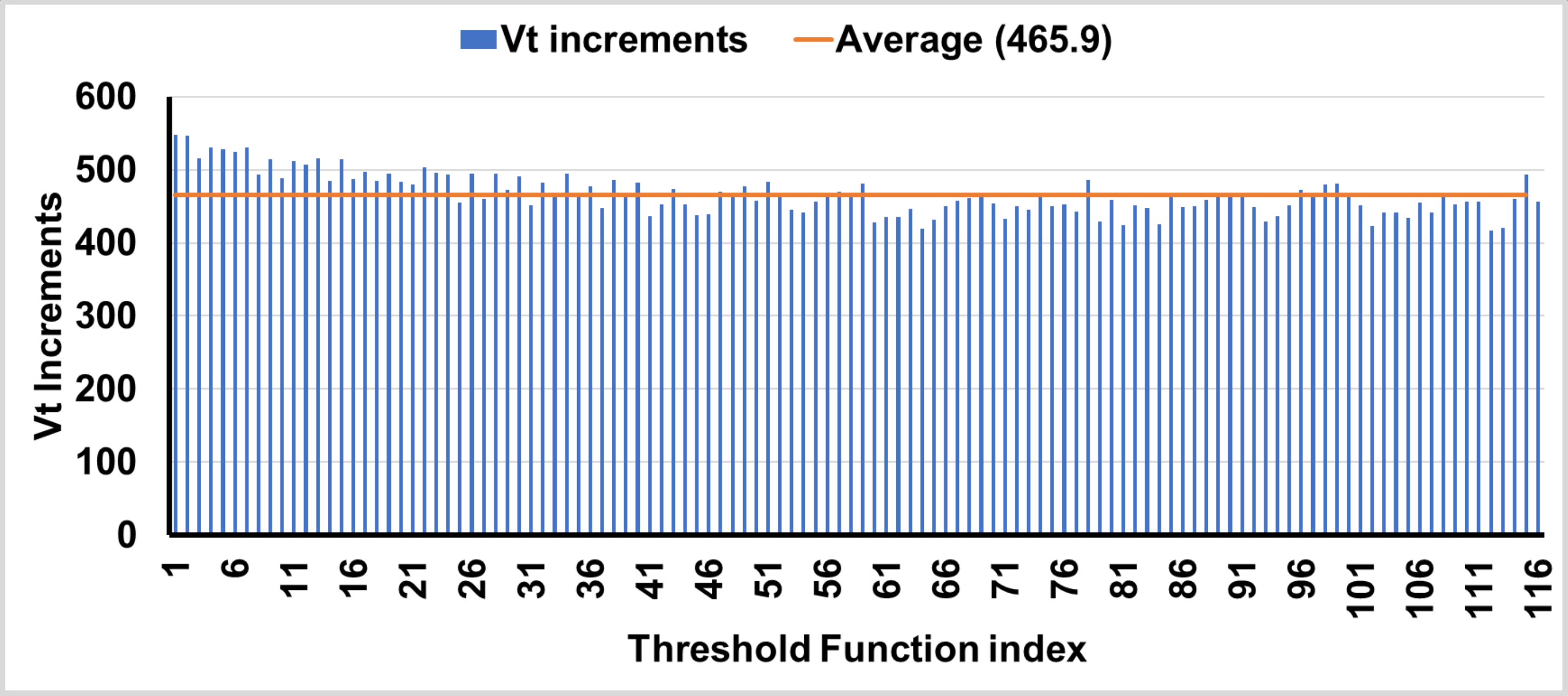}
		\caption{{\small Number of high voltage (HiV) pulses needed to program the FTL cells with all 117 threshold functions of up to 5 inputs}}
		\label{fig:ftl_pulse_requirement} 
	\end{figure}

	Figure \ref{fig:ftl_pulse_requirement} shows the number of high voltage pulses needed to program the 117 threshold functions. The number of high voltage pulses is estimated, assuming that each high voltage pulse would increment the threshold voltage of a flash transistor by 20mV. This assumption will vary across flash transistors. As shown in Figure \ref{fig:ftl_pulse_requirement}, the number of high voltage pulses needed to program a given FTL cell increases with an increase in the number of variables of the threshold function being implemented.
	
	\subsection{Experiments on Training for Robustness}
	\label{subsec:expResultsRobustness}
	
	In this section, we present the results of Algorithms $\text{mPLA}^{+}$, and $\text{mPLA}^{++}$ for training FTL cells taking into account parasitics and manufacturing variations. The test function$f_{35}(a,b,c,d,e): (\bm{W}, T) = [3, 3, 2, 1, 1; 8] =  ab(c + de)$ was chosen for this evaluation as this function generated the most number of error-types ($M_f$=61) out of all the 117 threshold functions when Monte Carlo simulations were run on 20K training samples. 
	
	The first experiment consisted of training $FTL_{35}$ using $\text{mPLA}^{+}$ for various values of the capacitances $C_1$ and $C_0$, and for each solution, extracting the delay values. The results for this experiment are as shown in Table \ref{cap_delay}. There are two important observations to be made here. First, even though the weights of the inputs $d$ and $e$ are equal, the corresponding flash transistors ($V_{4}$ and $V_{5}$) may be assigned different threshold voltages. This is because $\text{mPLA}^{+}$ compensated for the irregular layout parasitics of both the flash transistors using threshold voltages to realize equal weights. Second, the delay \textit{improves} with increasing robustness, as discussed earlier in Section \ref{robustness_training}. This is because the separation between the lines $G_R = G_L-\Delta_L$ and $G_R = G_L+\Delta_R$ increases with increase in $C_1$ and $C_0$. This increased separation results in a higher voltage difference at the inputs of the sense amplifier, which leads to a faster evaluation of the FTL cell.
	
	\begin{table}[ht]
		\centering
		
		\small
		\begin{tabular}{|l|l|l|}
			\hline
			$C_1,$   & Average Vt Values (V) & Delay\\
			$C_0$ & ($V_1, V_2, V_3, V_4, V_5; V_{l0}, V_{r0}$)  & (ps) \\  \hline
			0                                                          & 0.64, 0.64, 0.66,   0.70, 0.72; 1.00, 0.58                                                                  & 224                                                                       \\ \hline
			0.01                                                       & 0.60, 0.60, 0.64, 0.68, 0.70; 1.00, 0.50                                                                    & 178                                                                   \\ \hline
			0.02                                                       & 0.60, 0.60, 0.64, 0.68, 0.70; 1.00, 0.50                                                                    & 178                                                                    \\ \hline
			0.05                                                       & 0.60, 0.60, 0.64, 0.70, 0.70; 1.00, 0.50                                                                    & 172                                                                    \\ \hline
			0.1                                                        & 0.56, 0.56, 0.60, 0.66, 0.66; 1.00, 0.42                                                                    & 163                                                                   \\ \hline
			0.15                                                       & 0.52, 0.54, 0.58, 0.64, 0.64; 1.00, 0.34                                                                    & 154                                                                    \\ \hline
		\end{tabular}
		\caption{\label{cap_delay}\small Delay values of $FTL_{35}$ = [3, 3, 2, 1, 1; 8], trained for robustness using various capacitor values (fF).}
	\end{table}
	
	The second experiment was aimed at validating $\text{mPLA}^{++}$.  We used $f_{35}$ as a test function.  The first step is to create the \textit{database} $\{\bm{VT}^{++}(f_{35})\}$.  Algorithm $\text{mPLA}^{++}$ was given $f_{35}$ and $N_{MC} = 20K$ as inputs.  The erroneous instances were grouped into $M_{f_{35}} = 61$ error-types.   Algorithm $\text{mPLA}^{++}$  generated $\{\bm{VT}^{++}(f_{35})\} = \{\bm{VT}^+(f^e_{35,1}), \cdots, \bm{VT}^+(f^e_{35,61})\}$.  
	
	Next, 100K new MC instances were generated and programmed first with $\bm{VT}^{+}(f_{35})$. Among the erroneous instances, $99.96\%$ of them were one of 61 error-types that were previously found.  When each FTL cell in group $j, (1 \leq j \leq 61)$ was programmed with the threshold voltage set $\bm{VT}^{+}(f_{35,j})$, all the erroneous instances correctly computed $f_{35}$.  The remaining $.04\%$ of the 100K were correctly programmed by executing $\text{mPLA}^{0}$  directly to the chip, starting with $\bm{VT}^{+}(f_{35})$.  This required fewer than five iterations on the average for the instances.  Since $f_{35}$ had the most number of failure types, all of the other 117 functions, which exhibit fewer failure types, would be equally easy to program correctly in the presence of variations. Thus, all errors caused by process variations were corrected, with the vast majority requiring a single, precomputed VT set and a small fraction requiring on-chip programming. 
	
	\begin{table}[ht]
		\centering
		\small
		\begin{tabular}{|c|c|c|}
			\hline
			Stage                                                                          & Procedure & Yield (\%) \\ \hline
			\begin{tabular}[c]{@{}c@{}}Training   (20K instances)\\      $M_f$=61\end{tabular} & $\text{mPLA}^{++}$   & 100\%      \\ \hline
			\multirow{2}{*}{Testing (100K   instances)}                                    & $\text{mPLA}^{++}$    & 99.96\%    \\ \cline{2-3} 
			& $\text{mPLA}^0$ (On-chip) & 100\%      \\ \hline
		\end{tabular}
		\caption{\label{yield_table}\small Yield when $\text{mPLA}^{++}$ and $\text{mPLA}^0$ (on-chip) are used for programming instances of $FTL_{35}$ = [3, 3, 2, 1, 1; 8].}
		\vspace{-20pt}
	\end{table}
	
	\subsection{Robustness Against PVT Variations}
	
\begin{figure}[ht]
		\centerline{\includegraphics[width=\columnwidth]{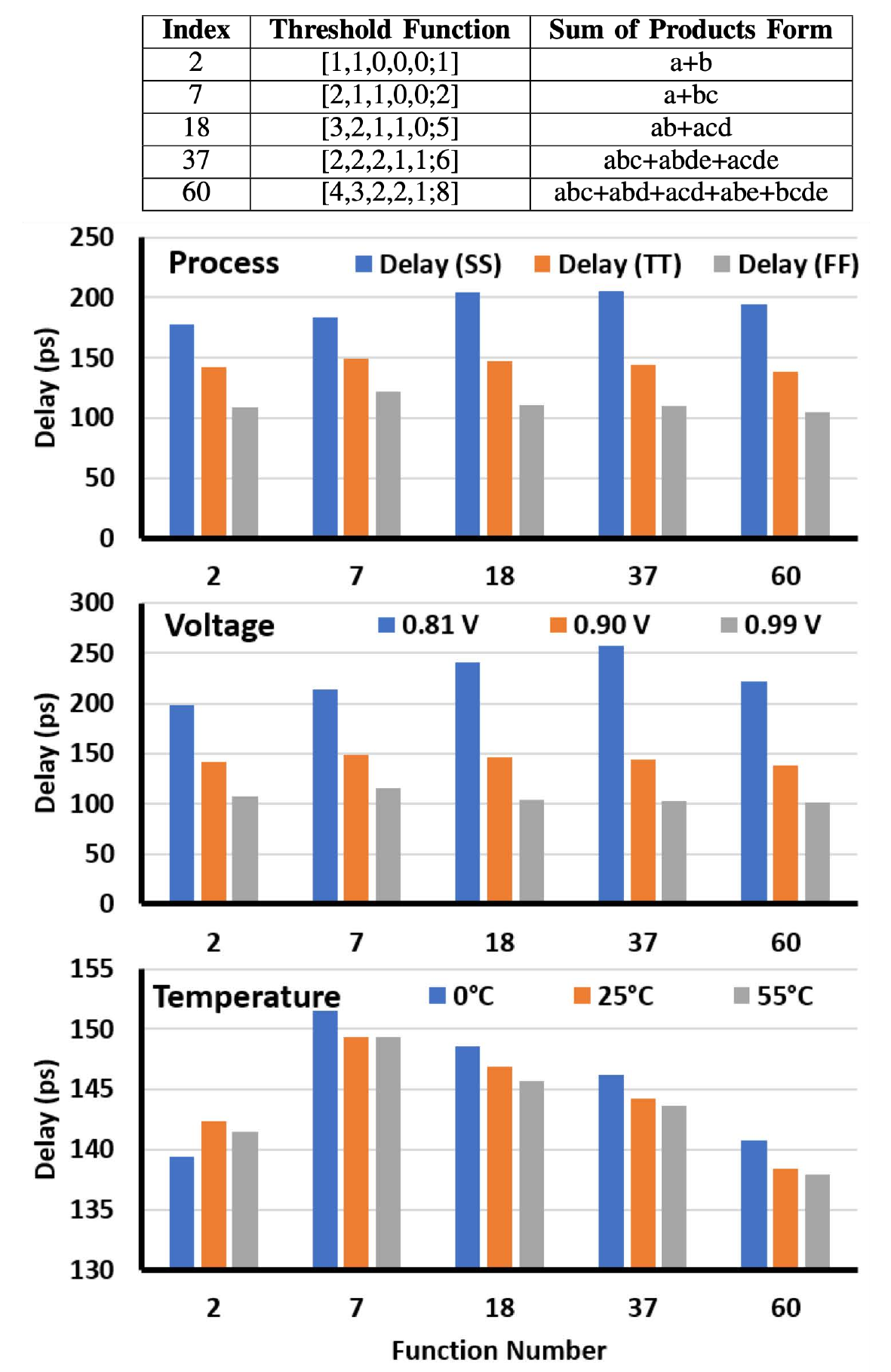}}
		\caption{\label{fig:temperature_delay}\small Delay of an FTL cell for threshold functions, with process (SS, TT, FF), voltage (0.81 V, 0.9 V, 0.99 V), and temperature (0$^{\circ}$C, 25$^{\circ}$C, 55$^{\circ}$C) variations.}
	\end{figure}

	Figure \ref{fig:temperature_delay} shows the delay variations in delay of five sample threshold functions~\cite{Threshold5List} w.r.t process, temperature and temperature variations.    As expected, FTL cells are slowest in the SS corner and fastest in the FF corner. Furthermore, as the process moves from the SS corner to the FF corner, the delay improves, as expected. When the voltage increases from 0.81 V to 0.99 V, the delay improves. The FTL cells were also tested for reliability for the consumer temperature range of 0$^{\circ}$C, 25$^{\circ}$C, and 55$^{\circ}$C. This result demonstrates that a $\bm{VT}^+(f)$ solution, generated using TT 0.9V 25$^{\circ}$C can reliably work with PVT variations. 
	
	
	\subsection{Robustness Against $VT$ Drift}
	
	Over the lifetime of an FTL cell, the charge stored in the gate of flash transistors eventually leaks into the channel due to the deterioration of thin oxide layer~\cite{Degraeve_Kaczer_Groeseneken_1999}, signal disturbances~\cite{Bersuker_Jeon_Huff_2001}, etc. This leakage effectively changes the $VT$ of the flash transistors. By extension, it also changes the weights programmed on the FTL cell. Table~\ref{Vt_drift} shows the effect of decreasing $VT$ on the threshold functions programmed on the FTL cells.  All 117 FTL cells operated correctly with a $VT$ drift of up to $5mV$.  Beyond $5mv$, some cells failed. However, after testing, their $VT$s can be reprogrammed to compensate for this drift. Furthermore, all the FTL cells that were selected by Genus when synthesizing ASIC designs (See Section \ref{subsec:asic_results}) operated correctly with $20 mV$ drift in $VT$.
	
	\begin{table}[ht]
		\centering
		\small
		\begin{tabular}{|c|c|}
			\hline
			\textbf{Vt Drift   (mV)} & \textbf{\% FTL cells operated correctly} \\ \hline
			1                        & 100                                           \\ \hline
			2                        & 100                                           \\ \hline
			5                        & 100                                           \\ \hline
			10                       & 96.55                                         \\ \hline
		\end{tabular}
		\caption{\small Robustness against $V_T$ drift for FTL cells programmed with all 117 threshold functions of up to 5 inputs.}
		\label{Vt_drift}
	\end{table}

	\subsection{Post-fabrication Timing Correction}
	
	The experiments described in Sections \ref{subsec:expResultsRobustness}, \ref{subsec:DelayDistributions} and \ref{subsec:VoltageScaling} demonstrate the flexibility of FTL due to the possibility of configuring its function after fabrication. This characteristic can also be used to correct timing errors. 
	
	\begin{figure}[ht!]
		\centering
		\includegraphics[scale=0.22]{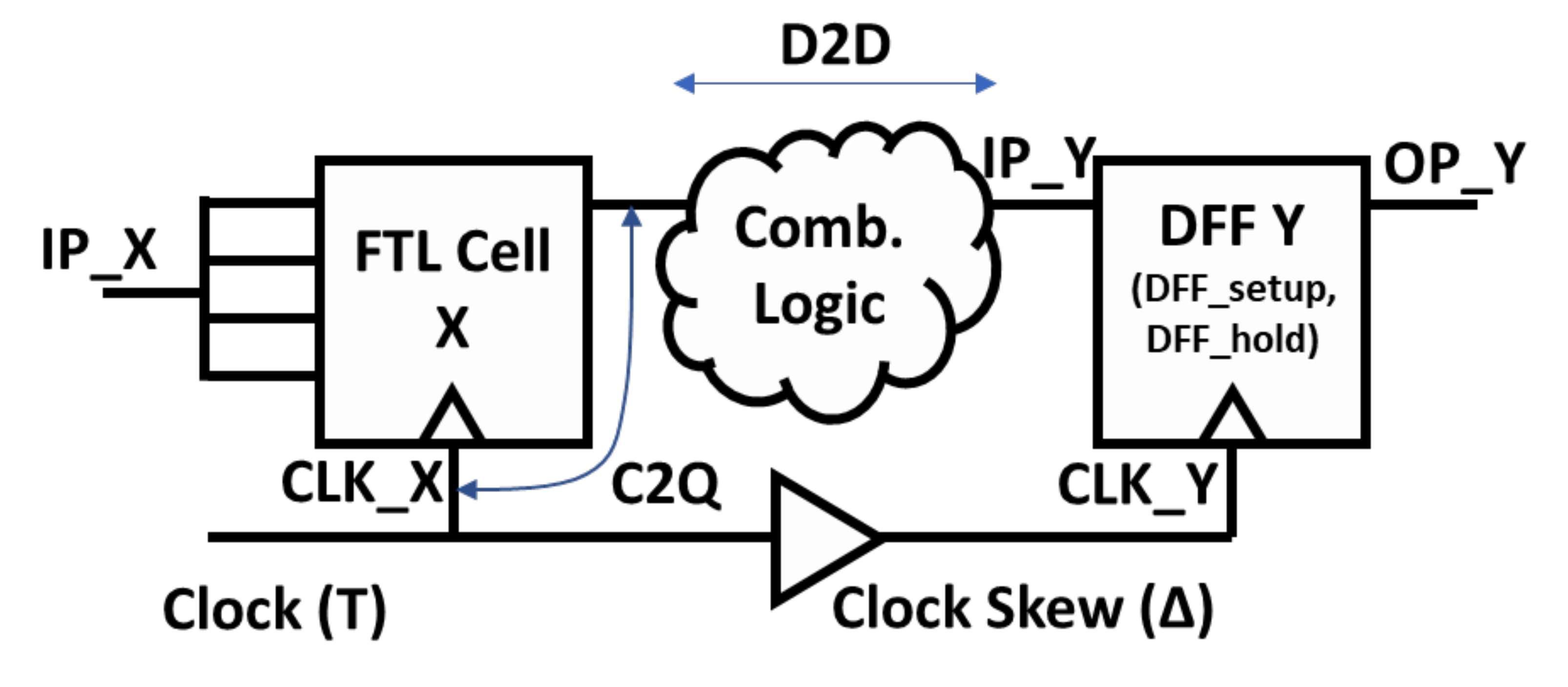}
		\caption{\small Datapath to Demonstrate Post-Fabrication Timing Corrections.}
		\label{fig:setup_hold_test_circuit}
	\end{figure}
	
	\begin{figure}[ht!]
		\centering
		\includegraphics[width=\columnwidth]{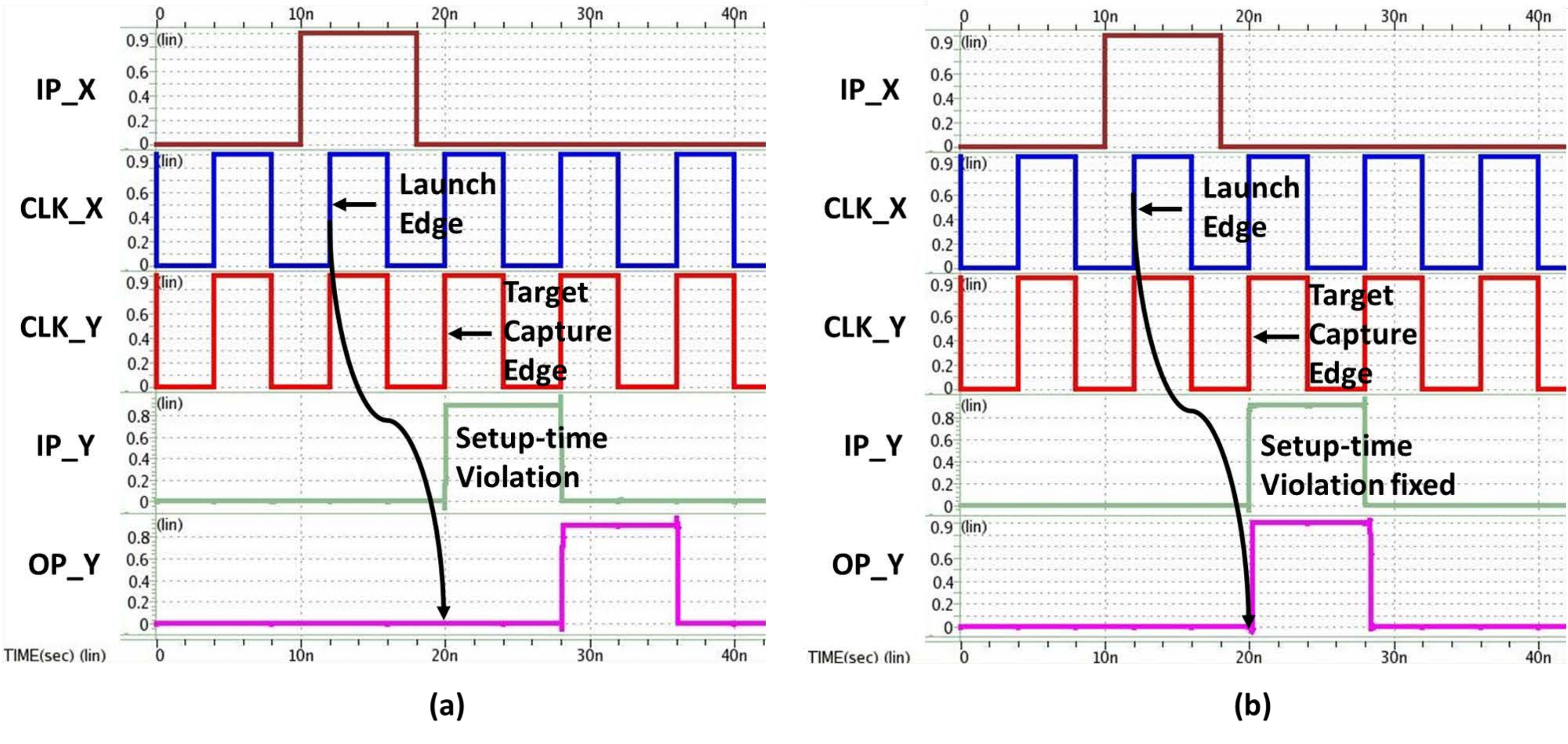}
		\caption{\small Post-fabrication setup-time correction using an FTL cell.}
		\label{setup_fix_waveform}
		\vspace{-10pt}
	\end{figure}
	
	\begin{figure}[ht!]
		\centering
		\includegraphics[width=\columnwidth]{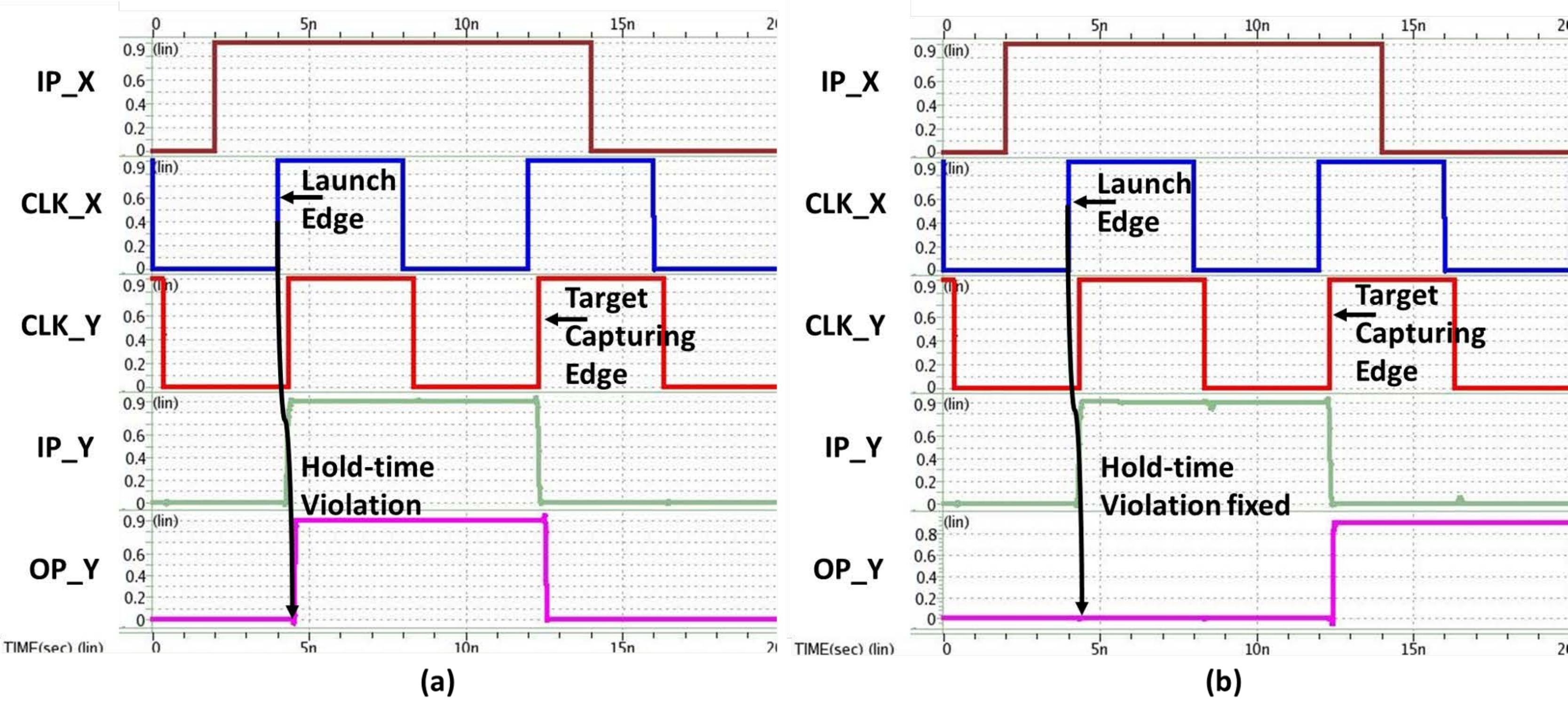}
		\caption{\small Post-fabrication hold-time correction using an FTL cell.}
		\label{hold_fix_waveform}
		\vspace{-10pt}
	\end{figure}
	~
	
	Figure~\ref{fig:setup_hold_test_circuit} shows a small datapath that was constructed to demonstrate how to correct setup-time and hold-time violations after fabrication in an FTL design. The datapath consists of clock-to-Q (C2Q) delay, combinational delay (D2D) and DFF specifications for setup ($DFF\_setup$) and hold ($DFF\_hold$) times. The clock is skewed by an appropriate amount $\Delta$, to generate either a setup-time or a hold-time violation. The violations are corrected by reprogramming the FTL cell to produce different C2Q values. 
	
	Figure \ref{setup_fix_waveform}(a) shows how the data launched from FTL X misses the target clock edge at DFF Y, thereby violating setup-time. Figure \ref{setup_fix_waveform}(b) then shows that decreasing the C2Q of FTL X fixes the setup-time violation. Similarly, Figure \ref{hold_fix_waveform}(a) shows how the data launched from FTL X is captured by the target clock edge at DFF Y one cycle early, thereby violating hold-time. Figure \ref{hold_fix_waveform}(b) then shows that increasing the C2Q of FTL X fixes the setup-time violation. Note that we can extend post-fabrication $V_T$ adjustment to also mitigate delay increases due to aging.
	
	\subsection{Delay Optimal Synthesis of ASICs with FTLs} 
	
	\label{subsec:asic_results}
	
	\begin{table*}[ht]
		\resizebox*{\textwidth}{!}{
			\begin{tabular}{|c|c|c|c|c|c|c|c|c|c|c|c|c|c|c|c|}
				\hline
				&  & \multicolumn{5}{c|}{Conventional} & \multicolumn{5}{c|}{FTL-integrated} & \multicolumn{3}{c|}{Improvements} &  \\ \hline
				Design & \begin{tabular}[c]{@{}c@{}}Freq. \\ (MHz)\end{tabular}    &  \begin{tabular}[c]{@{}c@{}}Std. \\ Cells\end{tabular} & DFF  & \begin{tabular}[c]{@{}c@{}}Area \\ ($\mu m^2$)\end{tabular} & \begin{tabular}[c]{@{}c@{}}Power \\ ($mW$)\end{tabular} & \begin{tabular}[c]{@{}c@{}}Wire-\\length\\ ($\mu m$)\end{tabular} & \begin{tabular}[c]{@{}c@{}}Std. \\ Cells\end{tabular} & DFF/FTL  & \begin{tabular}[c]{@{}c@{}}Area \\ ($\mu m^2$)\end{tabular} & \begin{tabular}[c]{@{}c@{}}Power\\  ($mW$)\end{tabular} & \begin{tabular}[c]{@{}c@{}}Wire-\\length \\ ($\mu m$)\end{tabular} & Area     & Power    & \begin{tabular}[c]{@{}c@{}}Wire-\\length \end{tabular} & \begin{tabular}[c]{@{}c@{}}Prog.\\ Time \\($\mu sec$) \end{tabular} \\ \hline \hline
				Mul & 417  &  19536      & 343  & 51855 & 6.00 & 118906 & 11493 & 272/71   & 32339 & 4.64 & 88592 & 37.6\%   & 22.7\%   & 25.5\%  & 204.3     \\ \hline
				Filter & 406  &53588    & 529  & 157482 & 36.26 & 436000 & 41711 & 281/248  & 107420 & 28.41 & 322400 & 31.8\%   & 21.6\%   & 26.1\%  & 585.8    \\ \hline
				FPU & 392 & 48992      & 1734 & 132655 & 27.82 & 484096 & 40937      & 1693/41  & 98879 & 24.14 & 406091 & 25.5\%   & 13.2\%   & 16.1\%  & 113.2    \\ \hline
				FFT  & 667  & 156242    & 9614 & 443356 & 100.14 & 1405565 & 140650     & 9286/328 & 368509  & 86.62  & 1199160 & 16.9\%   & 13.5\%   & 14.7\%  & 807.1    \\ \hline
				SHA & 308  & 33204      & 2161 & 109170 & 15.95 & 396267& 26511      & 2147/14  & 66001 & 13.23 & 343852 & 39.5\%   & 17.1\%   & 13.2\%   & 47.9   \\ \hline \hline
				Avg. &  &   &    & 139290     & 25         & 425689          &            &          & 96468       & 21         & 343504          & 30.7\%   & 17.7\%   & 19.3\%  & 351.7    \\ \hline
			\end{tabular}
		}
		\caption{\label{synthesis} Improvement in area, power, and wirelength improvement in ASICs with FTL integrated, over conventional ASICs, without trading off performance. Average improvements are calculated using the geometric mean.}
		\vspace{-10pt}
	\end{table*}
	
	In this section, we show how commercial design tools can accommodate FTL cells in synthesis, and placement and routing.  Five circuit blocks were synthesized using the 40nm TSMC standard cell library, which was augmented with FTLs to realize 117 positive forms of all 5-input threshold functions.  This was done by creating one cell and making 117 copies and then determining the $V_{T}$s of the flash transistors and signal assignments to realize each threshold function.  Then each FTL standard cell was characterized in the conventional way. Only the positive forms of the threshold functions were included in the library to keep the increase in the library size to a minimum (about 7\%) and to exploit the capability of Genus to recognize NPN equivalents of the cells (see below).
	
	The ASIC benchmarks are: 1) 32-bit Wallace multiplier (Mul), 2) 28-bit FIR filter (FIR), 3) 64-bit floating-point unit multiplier, 4) 16-bit Fast Fourier Transform (FFT), and 5) 512-bit Secure Hash Algorithm (SHA). Designs were synthesized using Cadence Genus and then placed and routed using Cadence Innovus. Standard cell libraries for FTL cells were characterized using Synopsys HSPICE and generated in Liberty format. Timing checks were performed using cross-corner analysis at \{SS, 125C, 0.81V\}, \{TT, 25C, 0.9V\} and \{FF, 0C, 0.99V\} corners.  After placement and routing, the \textit{select} cells and the FTL programming logic cells (see Figure~\ref{fig:flash_prog_arch} are \textit{paired}.  Then engineering change order (ECO) commands stitch the programming scan chain. Since the latter uses high voltage nets, shielding nets are added to protect neighboring nets from high voltage signals. Both versions of each ASIC were verified using Cadence Conformal.
	
	The results of synthesis and P\&R, summarized in Table~\ref{synthesis}, demonstrate significant improvements in the area (30.7\%), power (17.7\%), and wirelength (19.3\%) averaged over the designs.  These improvements include the \textit{overhead} of the programming infrastructure described in Section \ref{sec:ftl_asic_integration}, which was less than 5\% in the worst case. Note that these \textit{across the board} improvements were obtained under \textit{delay-optimal synthesis}.  This would not be the case for area-optimal synthesis. 
	
	Wherever it was beneficial to improve timing, Genus found and replaced \textit{threshold logic cones} (not necessarily maximal fanout-free cones) driving DFFs with the appropriate FTL cell.  This led to a reduction in the number of standard cells. It ranged from 10\% to 42\%.  There are two causes for this reduction. First is the absorption of part of the fanin cone that is a threshold function driving the DFF into the FTL.  This eliminates all those cells.  A second source is the reduction of the subcircuit (e.g. $C$ in Figure~\ref{subfigb:FTL_ASIC}) that \textit{feeds} the fanin cone. The significant speed advantage of the FTL cell creates large positive slack at the outputs of the \textit{feeder} subcircuit. Consequently, to meet timing, Genus re-synthesizes the feeder with slower logic. Standard logic primitives such as inverters, 2-input gates,  3-input gates, inverters, and even AOI/OAI gates are reduced and the number of complex cells increased, reducing the total cell count.  
	
	The last column of Table~\ref{synthesis} shows estimates of the time (i.e., number of pulses) required to program the FTL cells, which increases linearly with the number of FTLs. Although the actual programming time will depend on the technology, it is expected to be on the order of microseconds~\cite{Richter2014}.  
	
	Table \ref{syn_param} shows the run-time of Genus during synthesis, for all the ASIC designs. While the inclusion of all the 117 FTL cells increases the library size slightly (about 7\%), FTL cells allow \textit{faster timing closure} by generating positive slack. Table \ref{syn_param} also shows the peak memory usage of Genus during synthesis, for all the ASIC designs. The peak memory requirements are almost identical even after adding the 117 threshold functions in the library.

	\begin{table}[ht]
		\centering
		\small
		\begin{tabular}{|l|ll|ll|}
			\hline
			& \multicolumn{2}{c|}{Runtime(sec)}                  & \multicolumn{2}{c|}{Peak   memory  (MB)}           \\ \hline
			& \multicolumn{1}{l|}{Conv.} & FTL-integ. & \multicolumn{1}{l|}{Conv.} & FTL-integ. \\ \hline
			Multiplier & \multicolumn{1}{l|}{1451}         & 636            & \multicolumn{1}{l|}{1269}         & 1292           \\ \hline
			Filter     & \multicolumn{1}{l|}{2596}         & 2893           & \multicolumn{1}{l|}{1401}         & 1439           \\ \hline
			FPU        & \multicolumn{1}{l|}{2947}         & 2724           & \multicolumn{1}{l|}{1273}         & 1262           \\ \hline
			FFT        & \multicolumn{1}{l|}{3102}         & 2653           & \multicolumn{1}{l|}{1421}         & 1416           \\ \hline
			SHA        & \multicolumn{1}{l|}{1838}         & 1790           & \multicolumn{1}{l|}{1297}         & 1292           \\ \hline
		\end{tabular}
		\caption{\label{syn_param} {Runtime and Peak memory usage for the synthesis of ASIC designs.}}
	\end{table}

	\begin{table*}[t]
		\centering
		\resizebox{0.9\textwidth}{!}{  
			\small
			\begin{tabular}{|l|l|l|}
				\hline
				\begin{tabular}[c]{@{}l@{}}Threshold   \\ function\end{tabular} & Verilog description of NPN   equivalent & Synthesis   result \\ \hline
				\begin{tabular}[c]{@{}l@{}}ab+ace+ade\\ +bcd+acd\end{tabular}   & y \textless{}= ((4*a)   + (3*b) + (2*c) + (2*d) + (1*e)) \textgreater{}= 7 ? 1:0; & FTL\_93   (4,3,2,2,1;7)  \\ \cline{2-3} 
				{[}4,3,2,2,1;7{]} & y   \textless{}= ((4*(!a)) + (3*b) + (2*(!c)) + (2*d) + (1*e)) \textgreater{}= 7 ? 1:0;   & \begin{tabular}[c]{@{}l@{}}FTL\_93   (4,3,2,2,1;7)  and \\ two inverters for   "a" and "c"\end{tabular}        \\ \cline{2-3} 
				& y   \textless{}= !((4*(!b)) + (3*c) + (2*(!d)) + (2*a) + (1*e)) \textgreater{}= 7 ? 1:0;  & \begin{tabular}[c]{@{}l@{}}FTL\_94   (4,3,2,2,1;6)  and \\ three inverters for   "a", "c" and "e"\end{tabular} \\ \hline
			\end{tabular}
		}
		\caption{\label{syn_table}{Detection of NPN equivalents of threshold functions using a library of 117 5-input FTL cells.}}
	\end{table*}
	
	To demonstrate that Genus can recognize NPN equivalences of positive-form threshold functions, we selected a number of threshold functions and negated and permuted their inputs and negated their output.  Table~\ref{syn_table} shows the result of one of the more complex functions. The interpretation of Table~\ref{syn_table} is as follows. Consider the threshold function $ab + ace + ade + bcd + acd$.  The weight-threshold description is $[4,3,2,2,1;7] = 4a + 3b + 2c + 2c + d \geq 7$, which is an $FTL_{93}$.  When Genus found a sub-circuit with input negation, $\bar{a}b + \bar{a}\bar{c}e + \bar{a}de + bcd + a\bar{c}d$, , it replaced it with an $FTL_{93}$ with $\bar{a}$ and $\bar{c}$ driving inputs $a$ and $c$. The last row shows that Genus can detect output negation and maps it to a different cell $FTL_{94}$ whose positive form is $[4,3,2,2,1;6] = 4a + 3b + 2c + 2c + d \geq 6$. In each case, the synthesis tool detected the threshold functions and their NPN equivalents, and added inverters as necessary, without using any additional standard logic gates such as AND, OR, etc. 
	
\begin{figure}[h]
		\centering
		\includegraphics[width=\columnwidth]{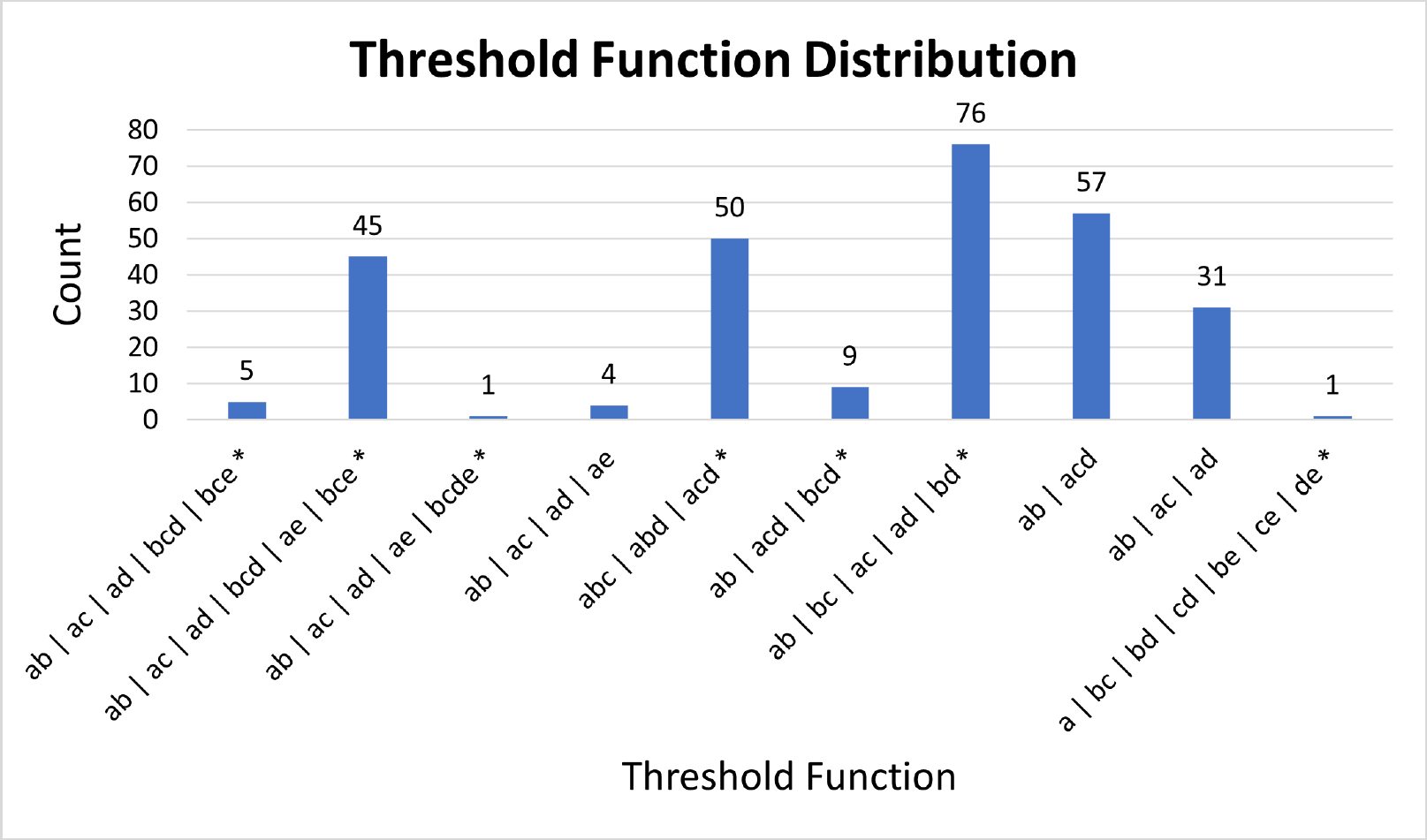}
		\caption{\label{fig:mul_dist}Distribution of threshold functions in 32-bit multiplier when synthesized using FTL cells with zero-delay zero-power.}
	\end{figure}

	The last experiment conducted was aimed at discovering what threshold functions would be detected if there were no area or delay constraints. Figure~\ref{fig:mul_dist} shows all possible threshold functions that could be detected in the 32-bit Wallace tree multiplier.  The multiplier has 343 DFFs. Excluding the 64 input DFFs, all 279 remaining DFFs and cones of logic feeding them were replaced by FTL cells, showing that complex multi-level logic circuits that are threshold functions frequently occur in logic circuits and synthesis tools can recognize them. 
	
	\vspace*{.5in}
	\section{Conclusion}
	\label{sec:Conclusions}
	
	In this paper we demonstrated that there could be substantial value in going beyond the traditional use of flash technology as memory and using it in CMOS logic.  Unlike the many emerging memory technologies, flash technology is mature and compatible with CMOS fabrication. Using flash transistors in conjunction with CMOS transistors, we developed a design of a binary neuron, referred to as FTL, that can realize a large number threshold functions in a single standard cell.  We demonstrated several novel features of an FTL cell: (1)~it is a \textit{configurable standard cell}, whose function can be configured after fabrication; (2) the configuration is achieved by conventional techniques of tuning the threshold voltages of flash transistors with high fidelity; (3) its design could be optimized to make it very robust in the presence of circuit parasitics and improving robustness also improves its performance; (4) the ability to tune its performance after fabrication provides a novel way to improve the yield in the presence of process variations and correct timing errors; (5)~ it was designed so that it can  automatically be embedded within ASICs using commercial CAD tools, and resulting in significantly improved area and power while still operating at the maximum possible frequency. 
	
	\nocite{dft,Neutzling_TCAD_2018}
	
	\bibliographystyle{unsrt}
	

	\newcommand{\bibspace}{-50pt}
	
	\begin{IEEEbiography}[{
			\includegraphics[width=1in,height=1.25in,clip,keepaspectratio]{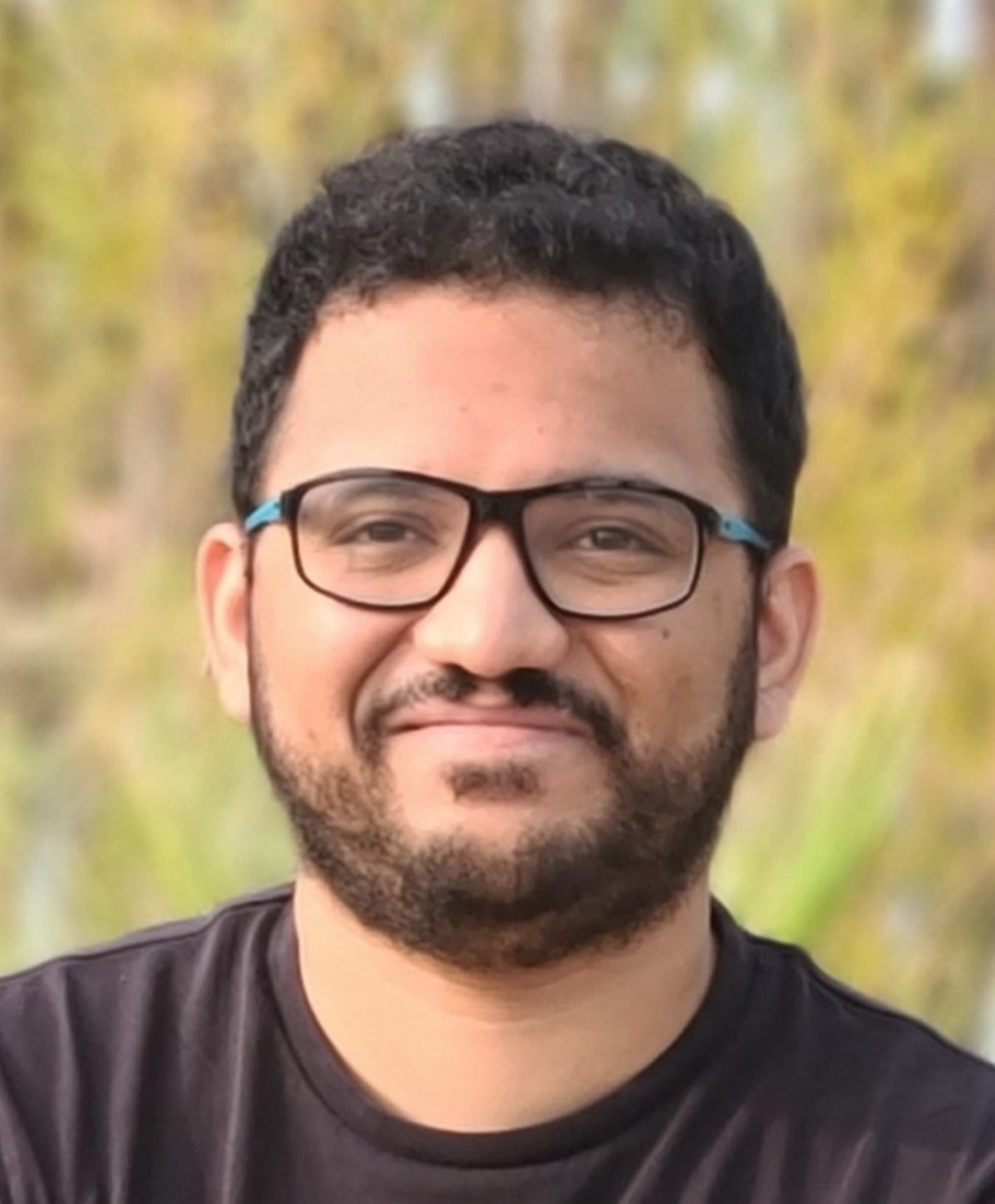}}]{Ankit Wagle} (M’17) received the B.S. degree in Electronics and Telecommunication from the University of Pune, Maharashtra, India, in 2013, and the M.S. degree in VLSI Design from Vellore Institute of Technology, Vellore, TN, India, in 2015. He spent his graduate research internships at Intel, Bangalore, KA, India in 2015 and Maxlinear, Carlsbad, CA, USA in 2017. He also worked with Open-Silicon, Bangalore, KA, India from 2015 to 2016. He is currently pursuing the Ph.D. degree with the School of Computing and Augmented Intelligence (SCAI), Arizona State University, Tempe, AZ, USA since 2016. His current research interests include new circuit architectures and design algorithms using threshold logic gates, and their applications to the design of energy efficient digital application-specified integrated circuit, field-programmable gate array, and neural network accelerators.
	\end{IEEEbiography}
	
	\vspace*{\bibspace}
	
	\begin{IEEEbiography}[{
			\includegraphics[width=1in,height=1.25in,clip,keepaspectratio]{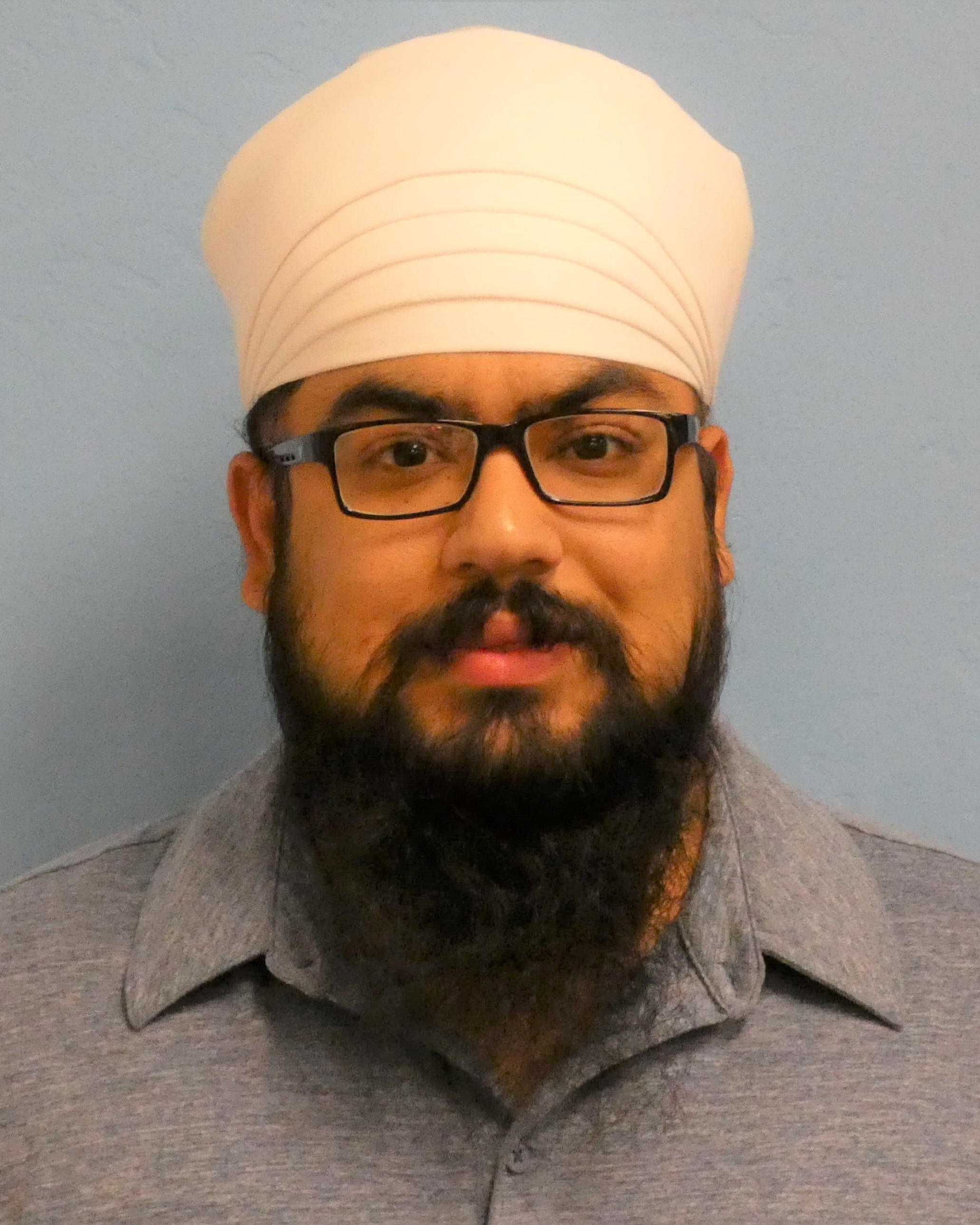}}]{Gian Singh} (M’19) received his B.Tech degree in Electronics and Communication Engineering from the National Institute of Technology (NIT-H), Hamirpur, India, in 2017. He worked as Project Associate at NIT-H under SMDP-C2SD project sponsored by the Govt. of India from 2017 to 2018. He started his Ph.D. degree at the School of Computing and Augmented Intelligence (SCAI), Arizona State University, Tempe, AZ, the USA in Fall 2018. He spent Fall’19 as an SoC Tech intern at Maxlinear Inc., Carlsbad, CA, USA and Summer’20 at Qualcomm Inc., San Jose, CA, USA as a Hardware Engineering intern. His current research interest includes the design of threshold logic gates, In-memory computing, near memory processing enabling high throughput and energy-efficient systems for data-intensive applications.
	\end{IEEEbiography}
	
	\vspace*{\bibspace}
	
	\begin{IEEEbiography}[{
			\includegraphics[width=1in,height=1.25in,clip,keepaspectratio]{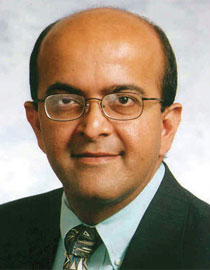}}]{Sunil Khatri} received the B.Tech. degree in electrical engineering from IIT Kanpur, Kanpur, India, the M.S. degree in electronics and communication engineering from The University of Texas at Austin, Austin, TX, USA, and the Ph.D. degree in electrical engineering and computer sciences from the University of California at Berkeley, Berkeley, CA, USA. He is currently a Professor of Electronics and Communication Engineering at Texas A\&M University, College Station, TX, USA. He has authored or coauthored more than 250 peer-reviewed publications. Among these papers, five received a best paper award, while six others received best paper nominations. He has coauthored nine research monographs and one edited research monograph, three book chapters, and 13 invited conference papers or workshop papers. He holds six U.S. patents. He was invited to serve as a Panelist at a conference seven times and have presented two conference tutorials. His current research interests include VLSI IC/system-on-a-chip design [including energy efficient design of custom ICs and field-programmable gate arrays (FPGAs), radiation and variation tolerant design, clocking], algorithm acceleration using hardware (FPGA as well as custom IC based) and software (uniprocessor and GPU based), and interdisciplinary extensions of these topics to other areas.
	\end{IEEEbiography}
	
	\vspace*{\bibspace}
	
	\begin{IEEEbiography}[{
			\includegraphics[width=1in,height=1.25in,clip,keepaspectratio]{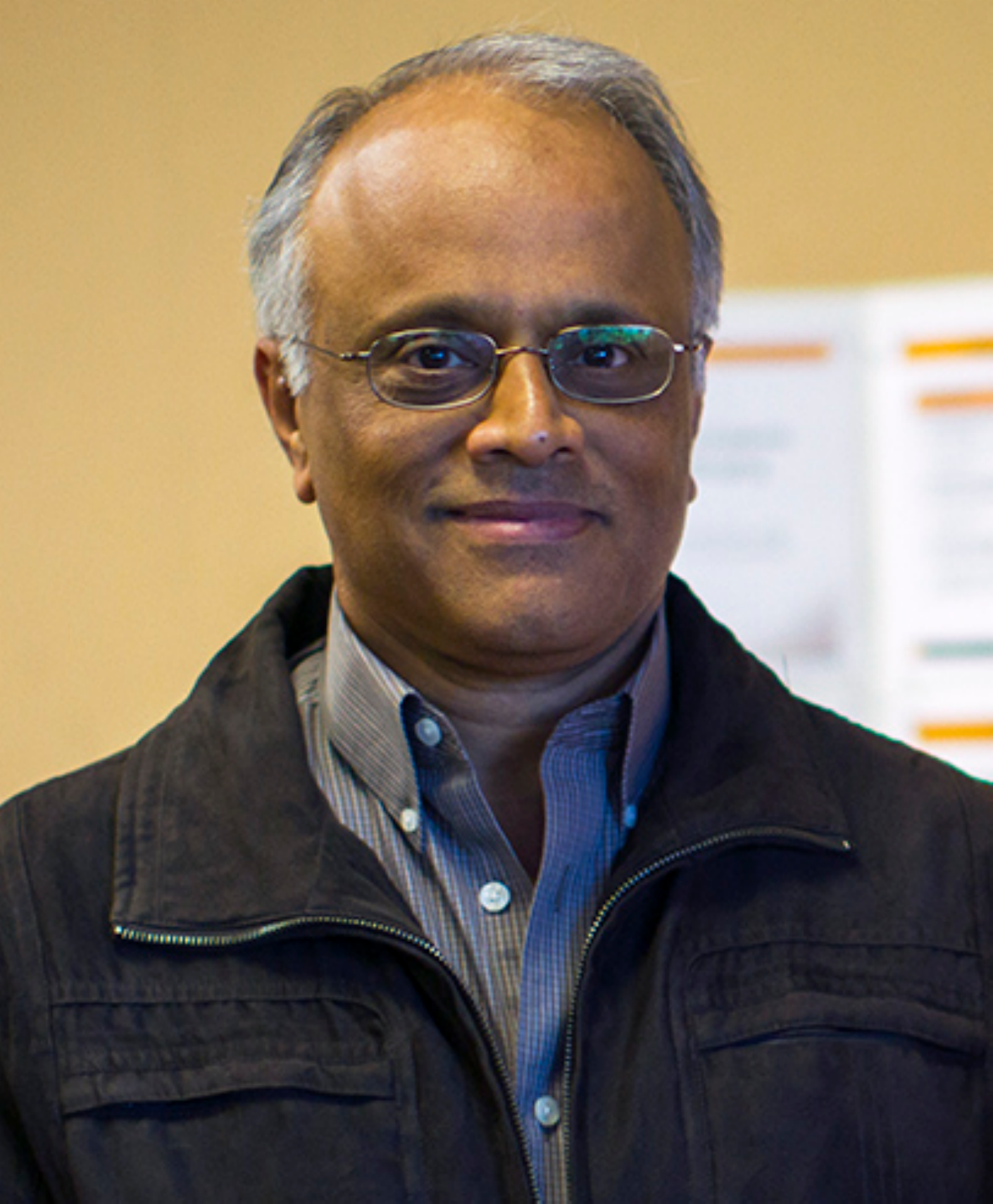}}]{Sarma Vrudhula} (M’85-SM’02-F'16) is a Professor of Computer Science and Engineering with Arizona State University, and the Director of the NSF I/UCRC Center for Embedded Systems.  He received the B.Math. degree from the University of Waterloo, Waterloo, ON, Canada, and the M.S.E.E. and Ph.D. degrees in electrical and computer engineering from the University of Southern California, Los Angeles, CA, USA.%
		His work spans several areas in design automation and computer aided design for digital integrated circuit and systems, focusing on low power circuit design, and energy management of circuits and systems. Specific topics include: energy optimization of battery powered computing systems, including smartphones, wireless sensor networks and IoT systems that rely on energy harvesting; system level dynamic power and thermal management of multicore processors and system-on-chip (SoC);  statistical methods for the analysis of process variations; statistical optimization of performance, power and leakage; new circuit architectures of threshold logic circuits for the design of ASICs and FPGAs.
	\end{IEEEbiography}
	
\end{document}